\documentclass{article}
\usepackage{amssymb}

\usepackage{graphicx}
\usepackage{amsmath}
\usepackage{a4wide}

\newtheorem{theorem}{Theorem}

\newtheorem{conclusion}{Conclusion}

\newtheorem{definition}{Definition}
\newtheorem{example}{Example}

\newtheorem{proposition}{Proposition}
\newtheorem{remark}{Remark}

\newenvironment{proof}[1][Proof]{\textbf{#1.} }{\ \rule{0.5em}{0.5em}}
\makeatletter
\def\@removefromreset#1#2{\let\@tempb\@elt
     \def\@tempa#1{@&#1}\expandafter\let\csname @*#1*\endcsname\@tempa
     \def\@elt##1{\expandafter\ifx\csname @*##1*\endcsname\@tempa\else
    \noexpand\@elt{##1}\fi}     \expandafter\edef\csname cl@#2\endcsname{\csname cl@#2\endcsname}     \let\@elt\@tempb
     \expandafter\let\csname @*#1*\endcsname\@undefined}

\@removefromreset{equation}{section}

\@removefromreset{theorem}{section}
\makeatother

\input{tcilatex}

\begin{document}

\title{General Probabilistic Framework of Randomness}
\author{Elena R. Loubenets \\
MaPhySto\thanks{%
A Network in Mathematical Physics and Stochastics, funded by The Danish
National Research Foundation.}\\
Department of Mathematical Sciences, University of Aarhus, Denmark}
\maketitle

\begin{abstract}
We introduce a new mathematical framework for the probabilistic description
of an experiment upon a system of any type in terms of initial information
representing this system. Based on the notions of an information state, an
information state space and a generalized observable, this general framework
covers the description of a wide range of experimental situations including
those where, with respect to a system, an experiment is perturbing.

We prove that, to any experiment upon a system, there corresponds a unique
generalized observable on a system initial information state space, which
defines the probability distribution of outcomes under this experiment. We
specify the case where initial information on a system provides ''no
knowledge'' for the description of an experiment.

Incorporating in a uniform way the basic notions of conventional probability
theory and the non-commutativity aspects and the basic notions of quantum
measurement theory, our framework clarifies the principle difference between
Kolmogorov's model in probability theory and the statistical model of
quantum theory. Both models are included into our framework as particular
cases. We show that the phenomenon of ''reduction'' of a system initial
information state is inherent, in general, to any non-destructive experiment
and upon a system of any type.

Based on our general framework, we introduce the probabilistic model for the
description of non-destructive experiments upon a quantum system and prove
that positive bounded linear mappings on the Banach space of trace class
operators, arising in the description of experiments upon a quantum system,
are completely positive.
\end{abstract}

\tableofcontents

\section{Introduction}

The problem of the relation between the statistical model of quantum theory
and conventional probability theory is a point of intensive discussions,
beginning from von Neumann's axioms [1] in quantum measurement theory and
Kolmogorov's axioms [2] in the theory of probability.

In the physical literature on quantum physics one can find statements on the
peculiarities of ''quantum'' probabilities and ''quantum'' events. In the
mathematical physics literature, the structure of conventional probability
theory is often referred to as Kolmogorov's model or as classical
probability, and it is argued that the Kolmogorov model is embedded as a
particular case into the so-called ''non-commutative probability theory'' -
the algebraic framework based on the structure of the statistical model of
quantum theory.

However, since the algebraic framework does not inherit the structure of
probability theory and does not cover the description of all possible
general probabilistic situations, this framework cannot be considered to
represent an extension of conventional probability theory. Moreover, the
algebraic framework cannot also, in principle, incorporate the developments
of quantum measurement theory.

Many attempts have been also made to include the statistical model of
quantum theory into the formalism of conventional probability theory.
However, all these attempts cannot be constructive, since random variables
(classical observables) in Kolmogorov's model represent only non-perturbing
experiments and, in general, this is not true for an experiment upon a
quantum system.

In the present paper we formulate the basics of a new general framework for
the probabilistic description of an experiment upon a system in terms of
initial information representing this system. Our mathematical setting is
most general and covers, in particular, those probabilistic situations where
a system is represented initially by a set of maximally available ''bits''
of information and a probability distribution of possible ''bits''.

We introduce the notions of an information state, an information state space
and a generalized observable and prove that any experiment\ upon a system is
represented on an initial information state space of this system by a
uniquely defined generalized observable. We discuss the situation where an
initial information state space of a system provides ''no knowledge'' for
the description of an experiment.

Our framework incorporates in a uniform way the basic notions of
conventional probability theory, the non-commutativity aspects and the basic
notions of quantum measurement theory.\ This allows us to clarify the
principle difference between Kolmogorov's model and the statistical model of
quantum theory. Both models are included into our framework as particular
cases.

We introduce the concept of a complete information description of a
non-destructive experiment upon a system and show that the phenomenon of
''reduction'' of an information state is, in general, inherent to any
non-destructive experiment and upon a system of any type. In the most
general settings, this phenomenon is induced by a ''renormalization'' of the
information on a system, conditioned upon the recorded outcome under a
single trial, and by the ''dynamical'' change of a system information state
in the course of a perturbing experiment. A ''reduction'' of a mixed
Kolmogorov probability state occurs, in particular, even under a
non-perturbing classical measurement. Since, as we establish in this paper,
the probabilistic model of quantum theory represents a special model of our
general framework, the well-known von Neumann quantum ''state collapse'',
postulated in [1], and its further generalizations represent particular
cases of this general phenomenon.

In case where an information space of a system has a Banach space based
structure, we define, in the most general settings, the notions of a mean
information state, a conditional posterior mean information state and the
concept of a complete statistical description of a non-destructive
experiment. We show that, for the definite class of non-destructive
experiments upon a system of this type, the change from an initial mean
information state to a conditional posterior mean information state is given
by the notion of a mean information state instrument, which we introduce in
this paper.

We formulate the probabilistic model for the description of a
non-destructive experiment upon a quantum system. This model specifies not
only the probability distribution of outcomes but also the conditional
probability distribution of posterior pure quantum state outcomes following
a single experimental trial.

Based on our framework, we prove that positive bounded linear mappings on
the Banach space of trace class operators, arising in the description of
non-destructive experiments upon a quantum system, are completely positive.
We note that, under the operational approach to the description of quantum
measurements, the complete positivity is always introduced axiomatically,
rather than actually proved as in the present paper.

The basics of the quantum stochastic approach to the description of quantum
measurements, formulated in [17-19], correspond to the general probabilistic
framework, introduced in this paper

\section{Description of experiments}

Consider an experiment upon a system of any type. Let the experimental
situation\ be specified by a\emph{\ }''\textit{complex of conditions, which
allows of any number of repetitions'' }[2]\textit{\ }and,\textit{\ }under
each trial, let an experimental outcome $\omega $ be of any kind. We denote
by $\Omega $ the set of all outcomes $\omega ,$ and by $\mathcal{F}_{\Omega
} $ a set of subsets of $\Omega ,$ which includes $\emptyset $ and $\Omega $
and represents mathematically possible questions on an outcome $\omega ,$
being posed under this experiment. Namely, each subset $B\in \mathcal{F}%
_{\Omega }\ $represents the event that an outcome $\omega $ belongs to $B$
is recorded. The pair $(\Omega ,\mathcal{F}_{\Omega })$ is called an outcome
space.

Suppose that an experimental situation is such that, under numerous
identical trials, the limit of relative frequencies of the occurrence of an
event $\omega \in B\in \mathcal{F}_{\Omega }$ exists (up to a measurement
error) and defines a non-negative number $\mathrm{Prob}\{\omega \in B\}\leq
1,$ called the probability, or the chance, that, under a single trial of
this experiment, the event $\omega \in B$ occurs. In this case we say that
this experimental situation admits the probabilistic description and, for
specificity, call such experimental situations \textit{normal. }

For any $B\in \mathcal{F}_{\Omega },$ denote $\mathrm{P}(B):=\Pr \mathrm{ob}%
\{\omega \in B\}$. The family $\mathbb{P}=\{\mathrm{P}(B):B\in \mathcal{F}%
_{\Omega }\}$ of probabilities of all events is called an outcome
probability law of this experimental situation. Clearly, $\mathrm{P}%
(\emptyset )=0,$ $\mathrm{P}(\Omega )=1,$ and $\mathrm{P}(B\cup B^{\prime })=%
\mathrm{P}(B)+\mathrm{P}(B^{\prime }),$ whenever $B\cap B^{\prime
}=\emptyset .$

According to Kolmogorov's axioms [2] in the theory of probability:

\noindent (i) $\mathcal{F}_{\Omega }$ is a\textit{\ }$\sigma $\textit{-}%
algebra on $\Omega $, so that $(\Omega ,\mathcal{F}_{\Omega })$ is a
measurable space;

\noindent (ii) an outcome probability law $\mathbb{P}$ is represented by a
normalized $\sigma $-additive positive real valued measure 
\begin{equation*}
\mu :\mathcal{F}_{\Omega }\rightarrow \lbrack 0,1],\text{ \ \ }\mu (\Omega
)=1,
\end{equation*}
such that $\mathrm{P}(B)=\mu (B),$ $\forall B\in \mathcal{F}_{\Omega }.$

A triple $(\Omega ,\mathcal{F}_{\Omega },\mu )$ represents a positive
measure space\footnote{%
On notions of measure theory, see, for example, [22,23].}. In conventional
probability theory\footnote{%
On the notions of conventional probability theory, see, for example, [10].}
a normalized $\sigma $-additive positive real valued measure $\mu $ and a
positive measure space $(\Omega ,\mathcal{F}_{\Omega },\mu )$ are called a
probability measure and a probability space, respectively.

Otherwise expressed:

\textit{To any normal experimental situation upon a system $\mathcal{S}$ of
any type, there corresponds the unique probability space }$(\Omega ,\mathcal{%
F}_{\Omega },\mu ),$ \textit{where the measurable space} $(\Omega ,\mathcal{F%
}_{\Omega })$ \textit{represents a space of outcomes and the normalized }$%
\sigma $\textit{-additive positive real valued measure }$\mu $ \textit{%
represents an outcome probability law }$\mathbb{P}$ \textit{of this
experimental situation.}

The above axioms are crucial and, to our knowledge, are valid for all
models, introduced in different concrete sciences, to describe normal
experimental situations.

These axioms are valid, in particular, in the quantum case where, under any
generalized quantum measurement with outcomes in a measurable space $(\Omega
,\mathcal{F}_{\Omega }),$ the outcome probability law is given\footnote{%
On the main notions of quantum measurement theory, see, for example,
[8,9,11-21].} by a normalized $\sigma $-additive positive real valued
measure 
\begin{equation*}
\mu (B)=\mathrm{tr}\{\rho M(B)\},\ \ \ \ \forall B\in \mathcal{F}_{\Omega },
\end{equation*}
expressed via a density operator $\rho ,$ \textrm{tr}$\{\rho \}=1,$ on a
separable complex Hilbert space $\mathcal{H}$ and a normalized $\sigma $%
-additive measure $M(\cdot )$ on $(\Omega ,\mathcal{F}_{\Omega }),$ with
values $M(B),$ $B\in \mathcal{F}_{\Omega }$, $M(\Omega )=I_{\mathcal{H}},$
that are non-negative bounded linear operators on $\mathcal{H}.$

\section{Probabilistic framework}

In this section we introduce, in the most general settings, the
representation of the outcome probability law of a normal experimental
situation upon a system in terms of information representing this system
before an experiment. This allows to formalize the probabilistic description
of all experimental situations upon a system, in particular, those on which
the initial information on a system provides ''no knowledge'' and also those
which perturb a system.

All experimental situations, discussed in this paper, are hypothetical.

\subsection{General settings}

Let $\mathcal{S}$ be a system of any type. In the most general settings, we
express the information, representing $\mathcal{S}$, by a positive measure
space 
\begin{equation*}
(\Theta ,\mathcal{F}_{\Theta },\pi )
\end{equation*}
where $\Theta $ is a set, $\mathcal{F}_{\Theta }$ is a $\sigma $-algebra\ of
subsets of $\Theta $ and $\pi $ is a normalized $\sigma $-additive positive
real valued measure on a measurable space $(\Theta ,\mathcal{F}_{\Theta })$.
The mathematical structure of a measurable space $(\Theta ,\mathcal{F}%
_{\Theta })$ is not specified. In particular, we do not, in general, presume
any linear or convex linear structure of a set $\Theta .$

This mathematical setting is most general and covers, in particular, those
situations where each element of $\Theta $\ is interpreted as a maximally
available ''bit'' of information on a system and a measure $\pi $\
represents a probability distribution of possible\textit{\ }$\theta \in
\Theta .$

Consider the description of an experiment $\mathcal{E},$ with outcomes in a
measurable space $(\Omega ,\mathcal{F}_{\Omega }),$ upon a system $\mathcal{S%
}$. Let, before an experiment, a system $\mathcal{S}$ be described by any of
the positive measure spaces in 
\begin{equation}
\{(\Theta ,\mathcal{F}_{\Theta },\pi ):\pi \in \mathcal{V}_{(\Theta ,%
\mathcal{F}_{\Theta })}\},  \label{1}
\end{equation}
where $\mathcal{V}_{(\Theta ,\mathcal{F}_{\Theta })}$ is the convex linear
set of all normalized $\sigma $-additive positive real valued measures on $%
(\Theta ,\mathcal{F}_{\Theta }).$

For any $\pi \in \mathcal{V}_{(\Theta ,\mathcal{F}_{\Theta })},$ denote by $%
\mathcal{E}+\mathcal{S}(\pi )$ the experimental situation where $\mathcal{E}$
is carried out upon $\mathcal{S}$, represented initially by $(\Theta ,%
\mathcal{F}_{\Theta },\pi ).$ If all experimental situations $\mathcal{E}+%
\mathcal{S}(\pi ),$ $\pi \in \mathcal{V}_{(\Theta ,\mathcal{F}_{\Theta })},$
are normal we call the experiment $\mathcal{E}$ upon $\mathcal{S}$ normal.
We consider further the description of only normal experiments $\mathcal{E}$
upon $\mathcal{S}$ and, therefore, suppress the term ''normal''.

According to the consideration in section 2, let a normalized $\sigma $%
-additive positive real valued measure 
\begin{equation*}
\mu _{\mathcal{E}}(\cdot ;(\Theta ,\mathcal{F}_{\Theta },\pi )):\mathcal{F}%
_{\Omega }\rightarrow \lbrack 0,1]
\end{equation*}
represent on $(\Omega ,\mathcal{F}_{\Omega })$ the outcome probability law
of an experimental situation $\mathcal{E}+\mathcal{S}(\pi ),$ $\pi \in 
\mathcal{V}_{(\Theta ,\mathcal{F}_{\Theta })}.$ For short, we further use
the notation $\mu _{\mathcal{E}}(\cdot ;\pi ):=\mu _{\mathcal{E}}(\cdot
;(\Theta ,\mathcal{F}_{\Theta },\pi ))$.

The mapping 
\begin{equation}
\mu _{\mathcal{E}}(\cdot ;\cdot ):\mathcal{F}_{\Omega }\times \mathcal{V}%
_{(\Theta ,\mathcal{F}_{\Theta })}\rightarrow \lbrack 0,1]  \label{2}
\end{equation}
describes all experimental situations $\mathcal{E}+\mathcal{S(}\pi ),$ $\pi
\in \mathcal{V}_{(\Theta ,\mathcal{F}_{\Theta })},$ that is, an experiment $%
\mathcal{E}$ upon a system $\mathcal{S}.$

Thus, \textit{to an experiment }$\mathcal{E}$\textit{\ upon a system }$%
\mathcal{S}$\textit{, represented initially by the information, expressed by 
}(\ref{1})\textit{, there corresponds the unique mapping (}\ref{2}).

However, the converse statement is not true and the same mapping (\ref{2})
may correspond to a variety of experiments upon $\mathcal{S}.$

In general, the initial information on $\mathcal{S}$ may be such that 
\begin{equation}
\mu _{\mathcal{E}}(B;\pi )=\mu _{\mathcal{E}}(B),\text{ \ \ \ }\forall B\in 
\mathcal{F}_{\Omega },\text{ \ \ }\forall \pi \in \mathcal{V}_{(\Theta ,%
\mathcal{F}_{\Theta })},  \label{3}
\end{equation}
and this relation implies that the initial information on $\mathcal{S},$
represented by (\ref{1}), is not relevant for the description of an
experiment $\mathcal{E}$ upon $\mathcal{S}.$

In this case, we say that\textit{\ the initial information on} $\mathcal{S}$ 
\textit{provides} ''\textit{no knowledge}'' \textit{for the description of
an experiment }$\mathcal{E}$ \textit{upon} $\mathcal{S}.$

If, however, the initial information on $\mathcal{S}$ provides ''the
knowledge'' on an experiment\textit{\ }$\mathcal{E}$ upon $\mathcal{S}$
then, in general, the randomness may be caused by\footnote{%
On the discussion of possible types of uncertainties, see [7].}:

(i) \textit{the uncertainty, encoded in a measure space }$(\Theta ,\mathcal{F%
}_{\Theta },\pi ),$ \textit{where, in the most general settings, two
mathematical objects, elements of a set }$\Theta $ \textit{and a measure} $%
\pi $, \textit{may be responsible for this; }

(ii) \textit{by a probabilistic set-up of an experiment} $\mathcal{E}$ 
\textit{itself}.

\noindent Notice that in case where an experiment is carried out upon a
physical microsystem, even if all macroscopic parameters of the experimental
set-up are defined with certainty, this may not be true for parameters,
characterizing the microscopic environment of the experimental device. The
latter is due to the fact that, in the most general case, we can not specify
definitely either a physical state of this microscopic environment or its
interaction with the observed microsystem.

In view of the \textit{informational context of a probability distribution} $%
\pi \in \mathcal{V}_{(\Theta ,\mathcal{F}_{\Theta })}$ in a measure space $%
(\Theta ,\mathcal{F}_{\Theta },\pi ),$ representing the initial information
on a system $\mathcal{S},$ it is natural to assume:\medskip

\noindent \textbf{Convention}. \textit{For a system} $\mathcal{S}$, \textit{%
any positive measure space} $(\Theta ,\mathcal{F}_{\Theta },\widetilde{\pi }%
) $ \textit{with a finite} $\sigma $-\textit{additive positive real valued
measure} $\widetilde{\pi }$ \textit{on} $(\Theta ,\mathcal{F}_{\Theta }),$ 
\textit{satisfying the relation} 
\begin{equation}
\widetilde{\pi }(\cdot )/\widetilde{\pi }(\Theta )=\pi (\cdot ),  \label{4}
\end{equation}
\textit{represents the same information on }$\mathcal{S}$ \textit{as }$%
(\Theta ,\mathcal{F}_{\Theta },\pi ).\medskip $

\noindent \textbf{Axiom (Statistical)}\footnote{%
This axiom \ is similar to the statistical axioms in [4,9,11,16],
introduced, however, under different settings.}. \textit{For an experiment} $%
\mathcal{E}$ \textit{with outcomes in} $(\Omega ,\mathcal{F}_{\Omega })$ 
\textit{upon a system} $\mathcal{S}$, \textit{the mapping} $\mu _{\mathcal{E}%
}(\cdot ;\cdot )$ \textit{is such that} 
\begin{equation}
\mu _{\mathcal{E}}(B;\pi )=\alpha _{1}\mu _{\mathcal{E}}(B;\pi _{1})+\alpha
_{2}\mu _{\mathcal{E}}(B;\pi _{2}),\text{ \ \ }\forall B\in \mathcal{F}%
_{\Omega },  \label{5}
\end{equation}
\textit{for any} \textit{measures }$\pi ,$ $\pi _{1},\pi _{2}\in \mathcal{V}%
._{(\Theta ,\mathcal{F}_{\Theta })},$ \textit{satisfying the relation} 
\begin{equation*}
\pi =\alpha _{1}\pi _{1}+\alpha _{2}\pi _{2},\text{ \ \ \ }\alpha
_{1},\alpha _{2}\geq 0,\text{ \ \ \ }\alpha _{1}+\alpha _{2}=1.
\end{equation*}

\begin{remark}
The statistical axiom is true even if the initial information provides ''%
\textit{no knowledge}'' for the description of an experiment (see (\ref{3})).
\end{remark}

Introduce the following notations.

(a) for any $\pi \in \mathcal{V}_{(\Theta ,\mathcal{F}_{\Theta })},$ we
denote by $[\pi ]$ the equivalence class of all finite $\sigma $-additive
positive real valued measures $\widetilde{\pi }$ on $(\Theta ,\mathcal{F}%
_{\Theta })$ equivalent to $\pi ,$ due to (\ref{4}), and by 
\begin{equation}
(\Theta ,\mathcal{F}_{\Theta },[\pi ]):=\{(\Theta ,\mathcal{F}_{\Theta },%
\widetilde{\pi }):\widetilde{\pi }\in \lbrack \pi ]\}  \label{6}
\end{equation}
the set of all positive measure spaces, representing, due to the convention,
the same information on $\mathcal{S}$ as $(\Theta ,\mathcal{F}_{\Theta },\pi
)$;

(b) we denote by $[\mathcal{V}_{(\Theta ,\mathcal{F}_{\Theta })}]$ the set 
\begin{equation}
\lbrack \mathcal{V}_{(\Theta ,\mathcal{F}_{\Theta })}]:=\{[\pi ]:\pi \in 
\mathcal{V}_{(\Theta ,\mathcal{F}_{\Theta })}\}  \label{7}
\end{equation}
of all equivalence classes $[\pi ]$;

(c) for any measurable space $(\Lambda ,\mathcal{F}_{\Lambda }),$ we denote
by $\mathcal{J}_{(\Lambda ,\mathcal{F}_{\Lambda })}$ the linear space of all 
$\sigma $-additive bounded real valued measures on $(\Lambda ,\mathcal{F}%
_{\Lambda })$ and by $\mathcal{J}_{(\Lambda ,\mathcal{F}_{\Lambda
})}^{(+)}\subset \mathcal{J}_{(\Lambda ,\mathcal{F}_{\Lambda })}$ the set of
all finite $\sigma $-additive non-negative real valued measures on $(\Lambda
,\mathcal{F}_{\Lambda }).$ Endowed with the norm 
\begin{equation*}
||\nu ||_{\mathcal{J}_{(\Lambda ,\mathcal{F}_{\Lambda })}}=\sup_{B\in 
\mathcal{F}_{\Lambda }}|\nu (B)|,\text{ \ \ \ }\forall \nu \in \mathcal{J}%
_{(\Lambda ,\mathcal{F}_{\Lambda })},
\end{equation*}
the normed linear space $\mathcal{J}_{(\Lambda ,\mathcal{F}_{\Lambda })}$ is
Banach\footnote{%
See [22], section 3.7.4.}.\medskip

Due to the convention, the information on $\mathcal{S}$, represented by $%
(\Theta ,\mathcal{F}_{\Theta },\pi ),$ is equivalently described by any
element of the equivalence class $(\Theta ,\mathcal{F}_{\Theta },[\pi ]).$
The latter implies that the mapping (\ref{2}) can be uniquely extended to
all of $\mathcal{J}_{(\Theta ,\mathcal{F}_{\Theta })}^{(+)}\backslash \{0\},$
with the property 
\begin{equation*}
\mu _{\mathcal{E}}(\cdot ;\widetilde{\pi }_{1})=\mu _{\mathcal{E}}(\cdot ;%
\widetilde{\pi }_{2}),\text{ \ \ }\forall \widetilde{\pi }_{1},\widetilde{%
\pi }_{2}\in \lbrack \pi ],\text{ \ \ }\forall \pi \in \mathcal{V}_{(\Theta ,%
\mathcal{F}_{\Theta })}.
\end{equation*}
Here ''$0$'' denotes the zero valued measure on $(\Theta ,\mathcal{F}%
_{\Theta }).$

Hence, to an experiment $\mathcal{E},$ with outcomes in $(\Omega ,\mathcal{F}%
_{\Omega }),$ upon a system $\mathcal{S},$ there corresponds the unique
mapping 
\begin{equation}
\mu _{\mathcal{E}}(\cdot ;\cdot ):\mathcal{F}_{\Omega }\times (\mathcal{J}%
_{(\Theta ,\mathcal{F}_{\Theta })}^{(+)}\backslash \{0\})\rightarrow \lbrack
0,1],  \label{8}
\end{equation}
where, for each $\widetilde{\pi }\in \mathcal{J}_{(\Theta ,\mathcal{F}%
_{\Theta })}^{(+)}\backslash \{0\},$ the mapping $\mu _{\mathcal{E}}(\cdot ;%
\widetilde{\pi })$ is a normalized $\sigma $-additive positive real valued
measure on $(\Omega ,\mathcal{F}_{\Omega })$ and 
\begin{equation}
\mu _{\mathcal{E}}(\cdot ;\alpha \widetilde{\pi })=\mu _{\mathcal{E}}(\cdot ;%
\widetilde{\pi }),\text{ \ \ \ }\forall \alpha >0.  \label{9}
\end{equation}
Furthermore, due to the statistical axiom and (\ref{9}), the mapping $\mu _{%
\mathcal{E}}(\cdot ;\cdot )$ must satisfy the condition 
\begin{equation}
\mu _{\mathcal{E}}(B;\widetilde{\pi })=\alpha _{1}\mu _{\mathcal{E}}(B;%
\widetilde{\pi }_{1})+\alpha _{2}\mu _{\mathcal{E}}(B;\widetilde{\pi }_{2}),
\label{10}
\end{equation}
for any $\widetilde{\pi }\in \lbrack \pi ],$ $\widetilde{\pi }_{1}\in
\lbrack \pi _{1}],$ $\widetilde{\pi }_{2}\in \lbrack \pi _{2}],$ such that 
\begin{equation*}
\pi =\alpha _{1}\pi _{1}+\alpha _{2}\pi _{2},\text{ \ \ }\alpha _{1},\alpha
_{2}\geq 0,\text{ \ }\alpha _{1}+\alpha _{2}=1.
\end{equation*}

\subsection{Information state spaces}

Based on our considerations in section 3.1, we introduce new mathematical
notions and prove the corresponding statements.

\subsubsection{Information states}

Let $(\Theta ,\mathcal{F}_{\Theta },[\pi ])$ be an equivalence class,
defined by (\ref{6}).

\begin{definition}[Information state]
\textit{We call an equivalence class} $(\Theta ,\mathcal{F}_{\Theta },[\pi
]) $ \textit{an} \textbf{information state.}
\end{definition}

If a $\sigma $-algebra $\mathcal{F}_{\Theta }$ is trivial, that is, $%
\mathcal{F}_{\Theta }=\{\emptyset ,\Theta \},$ then there exists only one
measure $\pi _{0}\in \mathcal{V}_{(\Theta ,\mathcal{F}_{\Theta })},$ with $%
\pi _{0}(\Theta )=1,$ $\pi _{0}(\emptyset )=0.$ We call the corresponding
information state $(\Theta ,\mathcal{F}_{\Theta },[\pi _{0}])$ \textit{%
trivial}.

If a $\sigma $-algebra $\mathcal{F}_{\Theta }$ contains all atom subsets $%
\{\theta \}$ of $\Theta ,$ we call an information state \textit{pure} if $%
\pi $ is a Dirac measure on $(\Theta ,\mathcal{F}_{\Theta })$ and \textit{%
mixed}, otherwise.

We say that an information state $(\Theta ,\mathcal{F}_{\Theta },[\pi ])$ is 
\textit{induced }by an information state $(\Theta ^{\prime },\mathcal{F}%
_{\Theta ^{\prime }},[\pi ^{\prime }])$ if a measure $\pi $ is subordinated
to a measure $\pi ^{\prime },$ that is, 
\begin{equation}
\pi (F)=\int_{\Theta ^{\prime }}\Phi (F;\theta ^{\prime })\pi ^{\prime
}(d\theta ^{\prime }),\text{ \ \ }\forall F\in \mathcal{F}_{\Theta },
\label{11}
\end{equation}
where $\Phi (\cdot ;\cdot ):\mathcal{F}_{\Theta }\times \Theta ^{\prime
}\rightarrow \lbrack 0,1]$ is a mapping\footnote{%
Called in probability theory as a Markov kernel from $(\Theta ^{\prime },%
\mathcal{F}_{\Theta ^{\prime }})$ to $(\Theta ,\mathcal{F}_{\Theta }).$}
such that:\newline
(i) for each $\theta ^{\prime }\in \Theta ^{\prime },$ the mapping $\Phi
(\cdot ;\theta ^{\prime })$ is a normalized $\sigma $-additive positive real
valued measure on $(\Theta ,\mathcal{F}_{\Theta })$;\newline
(ii) for each $F\in \mathcal{F}_{\Theta },$ the function $\Phi (F;\cdot
):\Theta ^{\prime }\rightarrow \lbrack 0,1]$ is $\mathcal{F}_{\Theta
^{\prime }}$-measurable.

If, in particular, 
\begin{equation*}
\Phi (F;\theta ^{\prime })=\chi _{\phi ^{-1}(F)}(\theta ^{\prime }),\text{ \
\ \ }\forall \theta ^{\prime }\in \Theta ^{\prime },\text{ \ \ }\forall F\in 
\mathcal{F}_{\Theta },
\end{equation*}
where $\chi _{F^{\prime }}(\cdot )$ is an indicator function of a subset $%
F^{\prime }\in \mathcal{F}_{\Theta ^{\prime }},$ a function $\phi :\Theta
^{\prime }\rightarrow \Theta $ is $\mathcal{F}_{\Theta ^{\prime }}/\mathcal{F%
}_{\Theta }$ measurable and $\phi ^{-1}(F)\in \mathcal{F}_{\Theta ^{\prime
}} $ is the preimage of a subset $F\in \mathcal{F}_{\Theta }$, then 
\begin{equation*}
\pi (F)=\pi ^{\prime }(\phi ^{-1}(F)),\text{ \ \ \ }\forall F\in \mathcal{F}%
_{\Theta }.
\end{equation*}
In this case, we say that an information state $(\Theta ,\mathcal{F}_{\Theta
},[\pi ])$ is an $\phi $-\textit{image }of the information state $(\Theta
^{\prime },\mathcal{F}_{\Theta ^{\prime }},[\pi ^{\prime }])$ and,
respectively, $(\Theta ^{\prime },\mathcal{F}_{\Theta ^{\prime }},[\pi
^{\prime }])$ is an $\phi $-preimage of $(\Theta ,\mathcal{F}_{\Theta },[\pi
]).$ We denote this $\phi $-image subordination of measures and states by 
\begin{eqnarray}
\pi (F) &=&(\pi ^{\prime }\circ \phi ^{-1})(F):=\pi ^{\prime }(\phi
^{-1}(F)),\text{ \ \ \ }\forall F\in \mathcal{F}_{\Theta },  \label{12} \\
(\Theta ,\mathcal{F}_{\Theta },[\pi ]) &=&\phi \lbrack (\Theta ^{\prime },%
\mathcal{F}_{\Theta ^{\prime }},[\pi ^{\prime }])].  \notag
\end{eqnarray}

In view of the notation (\ref{7}), denote by 
\begin{equation}
(\Theta ,\mathcal{F}_{\Theta },[\mathcal{V}_{(\Theta ,\mathcal{F}_{\Theta
})}]):=\{(\Theta ,\mathcal{F}_{\Theta },[\pi ]):\pi \in \mathcal{V}_{(\Theta
,\mathcal{F}_{\Theta })}\}  \label{13}
\end{equation}
the set of all information states on $(\Theta ,\mathcal{F}_{\Theta })$.

\begin{definition}
\textit{We call a set} $(\Theta ,\mathcal{F}_{\Theta },[\mathcal{V}_{(\Theta
,\mathcal{F}_{\Theta })}])$ \textit{an} \textbf{information state space}.
\end{definition}

We say that an information state space $(\Theta ,\mathcal{F}_{\Theta },[%
\mathcal{V}_{(\Theta ,\mathcal{F}_{\Theta })}])$ is \textit{trivial} if $%
\mathcal{F}_{\Theta }$ is trivial. In this case, $(\Theta ,\mathcal{F}%
_{\Theta },[\mathcal{V}_{(\Theta ,\mathcal{F}_{\Theta })}])$ consists of
only one \textit{trivial} information state $(\Theta ,\mathcal{F}_{\Theta
},[\pi _{0}])$.

If each state in $(\Theta ,\mathcal{F}_{\Theta },[\mathcal{V}_{(\Theta ,%
\mathcal{F}_{\Theta })}])$ is induced by a state in $(\Theta ^{\prime },%
\mathcal{F}_{\Theta ^{\prime }},[\mathcal{V}_{(\Theta ^{\prime },\mathcal{F}%
_{\Theta ^{\prime }})}])$ then we say that an information state space $%
(\Theta ,\mathcal{F}_{\Theta },[\mathcal{V}_{(\Theta ,\mathcal{F}_{\Theta
})}])$ is \textit{induced} by $(\Theta ^{\prime },\mathcal{F}_{\Theta
^{\prime }},[\mathcal{V}_{(\Theta ^{\prime },\mathcal{F}_{\Theta ^{\prime
}})}])$.

In particular, if measurable spaces $(\Theta ,\mathcal{F}_{\Theta })$ and $%
(\Theta ^{\prime },\mathcal{F}_{\Theta ^{\prime }})$ are isomorphic, that
is, there exists a bijection $f:\Theta ^{\prime }\rightarrow \Theta $ such
that functions $f$ and $f^{-1}$ are, respectively, $\mathcal{F}_{\Theta
^{\prime }}/\mathcal{F}_{\Theta }$ and $\mathcal{F}_{\Theta }/\mathcal{F}%
_{\Theta ^{\prime }}$ measurable, then, to each state $(\Theta ^{\prime },%
\mathcal{F}_{\Theta ^{\prime }},[\pi ^{\prime }])$ in $(\Theta ^{\prime },%
\mathcal{F}_{\Theta ^{\prime }},[\mathcal{V}_{(\Theta ^{\prime },\mathcal{F}%
_{\Theta ^{\prime }})}]),$ there is put into one-to-one correspondence the $%
f $-image state $(\Theta ,\mathcal{F}_{\Theta },[\pi ])$ in $(\Theta ,%
\mathcal{F}_{\Theta },[\mathcal{V}_{(\Theta ,\mathcal{F}_{\Theta })}]),$
with $\pi =\pi ^{\prime }\circ f^{-1},$ and vice versa.

\subsubsection{Statistical information states}

Let $(\Theta ,\mathcal{F}_{\Theta },[\mathcal{V}_{(\Theta ,\mathcal{F}%
_{\Theta })}])$ be an information state space.

\begin{definition}
\textit{Let} $\frak{R}$ \textit{be a set. For a mapping }$\Phi :(\Theta ,%
\mathcal{F}_{\Theta },[\mathcal{V}_{(\Theta ,\mathcal{F}_{\Theta
})}])\rightarrow \frak{R},$ \textit{we call values} $\eta _{\Phi }=\Phi
((\Theta ,\mathcal{F}_{\Theta },[\pi ])),$ $\pi \in \mathcal{V}_{(\Theta ,%
\mathcal{F}_{\Theta })},$ \textit{as} $\Phi $\textit{-}\textbf{statistical
information states} \textit{on} $(\Theta ,\mathcal{F}_{\Theta },[\mathcal{V}%
_{(\Theta ,\mathcal{F}_{\Theta })}]).$
\end{definition}

On $(\Theta ,\mathcal{F}_{\Theta },[\mathcal{V}_{(\Theta ,\mathcal{F}%
_{\Theta })}])$ there exists a variety of different types of statistical
information states.

Consider, in particular, the situation where $\mathrm{V}$ is a Banach space
and a mapping $\varphi :\Theta \rightarrow \mathrm{V}$ is measurable with
respect to the $\sigma $-algebra $\mathcal{F}_{\Theta }$ and the $\sigma $%
-algebra $\mathcal{B}_{\mathrm{V}}$ of Borel subsets of \textrm{V},\textrm{\ 
}and also bounded, that is, there exists some $C>0$ such that $||\varphi
(\theta )||_{\mathrm{V}}\leq C,$ $\forall \theta \in \Theta .$

The mapping 
\begin{equation*}
(\Theta ,\mathcal{F}_{\Theta },[\pi ])\mapsto \eta _{mean(\varphi )}((\Theta
,\mathcal{F}_{\Theta },[\pi ])):=\int_{\Theta }\varphi (\theta )\pi (d\theta
)\in \mathrm{V},\text{ \ \ \ }\forall \pi \in \mathcal{V}_{(\Theta ,\mathcal{%
F}_{\Theta })},
\end{equation*}
is well defined and $||\eta _{mean(\varphi )}||_{\mathrm{V}}\leq C.$ For
short, we denote 
\begin{equation*}
\eta _{mean(\varphi )}(\pi ):=\eta _{mean(\varphi )}((\Theta ,\mathcal{F}%
_{\Theta },[\pi ]))
\end{equation*}
and refer to $\eta _{mean(\varphi )}(\pi )$\textit{\ }as\textit{\ }a\textit{%
\ }$\varphi $-\textit{mean information} \textit{state\ }on $(\Theta ,%
\mathcal{F}_{\Theta },[\mathcal{V}_{(\Theta ,\mathcal{F}_{\Theta })}]).$%
\textit{\ }

The set\textit{\ }$\frak{R}_{mean(\varphi )}\subset $\textrm{V} of all $%
\varphi $-mean information states is convex linear and bounded.\textit{\ }

Denote by $(\Theta _{\varphi },\mathcal{B}_{\Theta _{\varphi }})$ a
measurable space where $\Theta _{\varphi }=\varphi (\Theta )\subset \mathrm{V%
}$ and $\mathcal{B}_{\Theta _{\varphi }}$ is the trace on $\Theta _{\varphi
} $ of the Borel $\sigma $-algebra $\mathcal{B}_{\mathrm{V}}$ on \textrm{V}.
We have $||\theta _{\varphi }||_{\mathrm{V}}\leq C,$ $\forall \theta
_{\varphi }\in \Theta _{\varphi },$ and, for any $\pi \in \mathcal{V}%
_{(\Theta ,\mathcal{F}_{\Theta })},$ 
\begin{equation*}
\eta _{mean(\varphi )}(\pi )=\int_{\Theta _{\varphi }}\theta _{\varphi }\pi
_{\varphi }(d\theta _{\varphi }),
\end{equation*}
where $\pi _{\varphi }(F)=\pi (\varphi ^{-1}(F)),$ $\forall F\in \mathcal{B}%
_{\Theta _{\varphi }}.$

In view of this representation, we further restrict our consideration in
section 4.5 of the statistical description of experiments only to the case
where 
\begin{eqnarray}
\Theta &=&\{\theta \in \mathrm{V}:||\theta ||_{\mathrm{V}}\leq C,\text{ for
some }C>0\},  \label{14} \\
\mathcal{F}_{\Theta } &\supseteq &\mathcal{B}_{\Theta }  \notag
\end{eqnarray}
For this case we introduce the special type of statistical information
states.

\begin{definition}[Mean information state ]
\textit{For an information state space,\ }with \textit{a measurable space }$%
(\Theta ,\mathcal{F}_{\Theta })$ \textit{satisfying} (\ref{14}), \textit{we
call the values of the linear mapping} 
\begin{equation*}
\pi \mapsto \eta _{mean}(\pi )=\int_{\Theta }\theta \pi (d\theta )\in 
\mathrm{V,}\text{ \ \ \ }\forall \pi \in \mathcal{V}_{(\Theta ,\mathcal{F}%
_{\Theta })},
\end{equation*}
as \textbf{mean information states} on\textit{\ }$(\Theta ,\mathcal{F}%
_{\Theta },[\mathcal{V}_{(\Theta ,\mathcal{F}_{\Theta })}]).$
\end{definition}

We say that the convex linear set\textit{\ }$\frak{R}_{mean}(\Theta )$ of
all mean information states on\ $(\Theta ,\mathcal{F}_{\Theta },[\mathcal{V}%
_{(\Theta ,\mathcal{F}_{\Theta })}])$\textit{\ }represents a\textit{\ mean
information state space.}

We say that a mean information state is \textit{pure} if it is represented
by an extreme element of $\frak{R}_{mean}(\Theta )$ and \textit{mixed},
otherwise.

In case of (\ref{14}), all atom subsets $\{\theta \}$ of $\Theta $ belong to 
$\mathcal{F}_{\Theta }$, the set of all extreme elements of $\frak{R}%
_{mean}(\Theta )$ is included in $\Theta ,$ and to each pure mean
information state there corresponds a unique pure information state.
However, the converse is not, in general, true and a pure information state
may correspond to a mixed mean information state. Moreover, a mixed mean
information state can be, in general, induced by a variety of information
states.

\subsection{Positive mapping valued measures}

Let $(\Lambda ,\mathcal{F}_{\Lambda })$ and $(\Theta ,\mathcal{F}_{\Theta })$
be any measurable spaces. Consider a mapping 
\begin{equation}
(\widetilde{\mathcal{K}}(\cdot ))(\cdot ):\text{ \ \ }\mathcal{F}_{\Lambda
}\times \lbrack \mathcal{V}_{(\Theta ,\mathcal{F}_{\Theta })}]\rightarrow
\lbrack 0,1],  \label{15}
\end{equation}
such that, for each equivalence class $[\pi ]\in \lbrack \mathcal{V}%
_{(\Theta ,\mathcal{F}_{\Theta })}],$ the mapping 
\begin{equation*}
(\widetilde{\mathcal{K}}(\cdot ))([\pi ]):\mathcal{F}_{\Lambda }\rightarrow
\lbrack 0,1],\ \ \ \ \ \ (\widetilde{\mathcal{K}}(\Lambda ))([\pi ])=1,
\end{equation*}
is a normalized $\sigma $-additive positive real valued measure on $(\Lambda
,\mathcal{F}_{\Lambda })$.

For a set $X,$ denote by $\frak{B(}X)$ the Banach space of all bounded
complex valued functions $\Psi :X\rightarrow \mathbb{C}$. Recall that, in $%
\frak{B(}X),$ the norm is defined by 
\begin{equation*}
||\Psi ||_{_{\frak{B(}X)}}=\sup_{x\in X}|\Psi (x)|,\text{ \ \ \ }\forall
\Psi \in \frak{B}\frak{(}X).
\end{equation*}

Denote by $\frak{B}_{+}(X)\subset \frak{B}\frak{(}X)$ the set of all
non-negative real valued bounded functions on $X$. For any $\Psi _{1},\Psi
_{2}\in \frak{B}\frak{(}X),$ we write $\Psi _{1}\leq \Psi _{2}$ if $(\Psi
_{2}-\Psi _{1})\in \frak{B}_{+}\frak{(}X).$ Let $\mathcal{I}_{_{\frak{B(}%
X)}}\in \frak{B}_{+}\frak{(}X)$ be the positive function 
\begin{equation*}
\mathcal{I}_{_{\frak{B(}X)}}(x)=1,\text{ \ \ \ }\forall x\in X.
\end{equation*}

We can now say that, in (\ref{15}), the mapping $\widetilde{\mathcal{K}}$ is
a normalized $\sigma $-additive\footnote{%
Here the $\sigma $-additivity is understood in the following sense: for any $%
\pi \in \mathcal{V}_{(\Lambda ,\mathcal{F}_{\Lambda })},$ a normalized
positive real valued measure $(\widetilde{\mathcal{K}}(\cdot ))([\pi ])$ is $%
\sigma $-additive.} measure on $(\Lambda ,\mathcal{F}_{\Lambda })$ with
values $\widetilde{\mathcal{K}}(B),$ $B\in \mathcal{F}_{\Lambda },$ that are
non-negative real valued bounded functions on the set $[\mathcal{V}_{(\Theta
,\mathcal{F}_{\Theta })}],$ that is, 
\begin{equation*}
\widetilde{\mathcal{K}}(B)\in \frak{B}_{+}([\mathcal{V}_{(\Theta ,\mathcal{F}%
_{\Theta })}]),\text{ \ \ }\forall B\in \mathcal{F}_{\Lambda }.
\end{equation*}
There is the one-to-one correspondence between $\widetilde{\mathcal{K}}$ and
the mapping 
\begin{equation*}
\mathcal{K}(\cdot ;\cdot ):\mathcal{F}_{\Lambda }\times (\mathcal{J}%
_{(\Theta ,\mathcal{F}_{\Theta })}^{(+)})\backslash \{0\})\rightarrow
\lbrack 0,1],
\end{equation*}
defined by the relation 
\begin{equation*}
\mathcal{K}(\cdot ;\pi ^{\prime }\}:=(\widetilde{\mathcal{K}}(\cdot ))([\pi
]),\text{ \ \ }\forall \pi ^{\prime }\in \lbrack \pi ],\text{ \ }\forall \pi
\in \mathcal{V}_{(\Theta ,\mathcal{F}_{\Theta })}.
\end{equation*}

We formulate the following theorem.

\begin{theorem}
\textit{Let} $(\Lambda ,\mathcal{F}_{\Lambda })$\textit{\ and} $(\Theta ,%
\mathcal{F}_{\Theta })$ \textit{be any measurable spaces. To a mapping\ } 
\begin{equation*}
\widetilde{\mathcal{K}}:\mathcal{F}_{\Lambda }\rightarrow \frak{B}_{+}([%
\mathcal{V}_{(\Theta ,\mathcal{F}_{\Theta })}]),\text{ \ }\widetilde{%
\mathcal{K}}(\Lambda )=\mathcal{I}_{_{\frak{B(}[\mathcal{V}_{(\Theta ,%
\mathcal{F}_{\Theta })}])}}\text{,}
\end{equation*}
\textit{satisfying the conditions:}\newline
(\textbf{i})\textit{\ for} \textit{each} $\pi \in \mathcal{V}_{(\Theta ,%
\mathcal{F}_{\Theta })},$ \textit{the mapping} ($\widetilde{\mathcal{K}}%
(\cdot ))([\pi ]):\mathcal{F}_{\Lambda }\rightarrow \lbrack 0,1]$, $(%
\widetilde{\mathcal{K}}(\Lambda ))([\pi ])=1,$\textit{\ is a normalized\ }$%
\sigma $\textit{-additive positive real valued measure\ on} $(\Lambda ,%
\mathcal{F}_{\Lambda })$;\newline
(\textbf{ii}) $(\widetilde{\mathcal{K}}(B))([\pi ])=\alpha _{1}(\widetilde{%
\mathcal{K}}(B))([\pi _{1}])+\alpha _{2}(\widetilde{\mathcal{K}}(B))([\pi
_{2}]),$ $\forall B\in \mathcal{F}_{\Lambda }$\newline
\textit{for any} $\pi ,\pi _{1},\pi _{2}\in \mathcal{V}_{(\Theta ,\mathcal{F}%
_{\Theta })},$ $\pi =\alpha _{1}\pi _{1}+\alpha _{2}\pi _{2},$\textit{\ }$%
\alpha _{1},\alpha _{2}\geq 0,$ $\alpha _{1}+\alpha _{2}=1;\medskip \newline
$\textit{there exists a unique normalized }$\sigma $\textit{-additive measure%
} 
\begin{equation*}
\Pi :\mathcal{F}_{\Lambda }\rightarrow \frak{B}_{+}(\Theta ),\text{ \ \ }\Pi
(\Lambda )=\mathcal{I}_{_{\frak{B(}\Theta )}},
\end{equation*}
\textit{with }$\Pi (B):\Theta \rightarrow \lbrack 0,1]$ \textit{being\ }$%
\mathcal{F}_{\Theta }$-\textit{measurable for} \textit{each} $B\in \mathcal{F%
}_{\Lambda };$\textit{\newline
such that} 
\begin{equation*}
((\widetilde{\mathcal{K}}(B))([\pi ])=\int_{\Theta }(\Pi (B))(\theta )\pi
(d\theta ),
\end{equation*}
\textit{for all} $B\in \mathcal{F}_{\Lambda },$ $\pi \in \mathcal{V}%
_{(\Theta ,\mathcal{F}_{\Theta })}.$\medskip
\end{theorem}

\subsection{Generalized observables}

Introduce the following general concept.

\subsubsection{Definition, properties}

Let $(\Theta ,\mathcal{F}_{\Theta })$ and $(\Lambda ,\mathcal{F}_{\Lambda })$
be any measurable spaces.

\begin{definition}[Generalized observable]
\textit{We call a normalized }$\sigma $\textit{-additive measure} 
\begin{equation*}
\Pi :\text{ }\mathcal{F}_{\Lambda }\rightarrow \frak{B}_{+}(\Theta ),\text{
\ \ \ }\Pi (\Lambda )=\mathcal{I}_{_{\frak{B(}\Theta )}},
\end{equation*}
\textit{where, for each }$B\in \mathcal{F}_{\Lambda },$ \textit{the function}
$\Pi (B)$ \textit{is} $\mathcal{F}_{\Theta }$-\textit{measurable}, \textit{a}
\textbf{generalized observable, }\textit{with an outcome space\ }$(\Lambda ,%
\mathcal{F}_{\Lambda })$ \textit{and on} $(\Theta ,\mathcal{F}_{\Theta })$.
\end{definition}

In the terminology of conventional probability theory, for each $B\in 
\mathcal{F}_{\Lambda },$ the $\mathcal{F}_{\Theta }$-measurable function $%
\Pi (B):\Theta \rightarrow \lbrack 0,1]$ is a random variable on $(\Theta ,%
\mathcal{F}_{\Theta }).$

The set of all generalized observables on $(\Theta ,\mathcal{F}_{\Theta })$,
with an outcome space $(\Lambda ,\mathcal{F}_{\Lambda }),$ is convex linear.
For any generalized observable $\Pi ,$ we have $\Pi (\emptyset )=0$ and 
\begin{equation*}
0\leq \Pi (B_{1})\leq \Pi (B_{2})\leq \mathcal{I}_{_{\frak{B(}\Theta )}},
\end{equation*}
for any $B_{1}\subseteq B_{2}\subseteq \Lambda .$

Since an information state space $(\Theta ,\mathcal{F}_{\Theta },[\mathcal{V}%
_{(\Theta ,\mathcal{F}_{\Theta })}])$ is defined uniquely by a measurable
space $(\Theta ,\mathcal{F}_{\Theta }),$ we further equivalently refer to $%
\Pi $ as a generalized observable on an information state space $(\Theta ,%
\mathcal{F}_{\Theta },[\mathcal{V}_{(\Theta ,\mathcal{F}_{\Theta })}]).$

\begin{definition}
\textit{We call a generalized observable }$\Pi $\textit{\ on}\textbf{\ }$%
(\Theta ,\mathcal{F}_{\Theta })$\textbf{\ trivial} \textit{if} \textit{it
has the form} 
\begin{equation*}
\Pi (B)=\nu (B)\mathcal{I}_{_{\frak{B(}\Theta )}},\text{ \ \ \ }\forall B\in 
\mathcal{F}_{\Lambda },
\end{equation*}
with some $\nu \in \mathcal{V}_{(\Lambda ,\mathcal{F}_{\Lambda })}.$
\end{definition}

If a $\sigma $-algebra $\mathcal{F}_{\Theta }$ is trivial then any $\mathcal{%
F}_{\Theta }$-measurable function $\Theta \rightarrow \lbrack 0,1]$ is
constant and, in this case, on $(\Theta ,\mathcal{F}_{\Theta })$ there exist
only trivial generalized observables.

If $\mathcal{F}_{\Lambda }$ is trivial then, in $\mathcal{V}_{(\Lambda ,%
\mathcal{F}_{\Lambda })},$ there is only measure $\pi _{0},$ with $\pi
_{0}(\Lambda )=1,$ $\pi _{0}(\emptyset )=0.$ In this case, on $(\Theta ,%
\mathcal{F}_{\Theta })$ there exists only one generalized observable $\Pi
_{0}=\pi _{0}\mathcal{I}_{_{\frak{B(}\Theta )}}$ with the outcome space ($%
\Lambda ,\mathcal{F}_{\Lambda }),$ and it is also trivial.

\begin{remark}
Due to definition 5, to any mapping $\widetilde{\mathcal{K}},$ specified in
theorem 1, there corresponds a unique\textit{\ }generalized observable%
\textit{\ }$\Pi $ on\textit{\ }$(\Theta ,\mathcal{F}_{\Theta }).$ If $%
\widetilde{\mathcal{K}}$ has the form 
\begin{equation*}
(\widetilde{\mathcal{K}}(\cdot ))([\pi ])=k(\cdot ),\text{ \ \ \ \ }\forall
\pi \in \mathcal{V}_{(\Theta ,\mathcal{F}_{\Theta })},
\end{equation*}
with $k\in \mathcal{V}_{(\Lambda ,\mathcal{F}_{\Lambda })}$, then the
corresponding generalized observable $\Pi $ is given by $\Pi (B)=k(B)%
\mathcal{I}_{_{\frak{B(}\Theta )}},$ $\forall B\in \mathcal{F}_{\Lambda },$
and is trivial.
\end{remark}

Denote by $\mathcal{G}_{(\Theta ,\mathcal{F}_{\Theta })}$ the set of all
non-trivial generalized observables on an information state space $(\Theta ,%
\mathcal{F}_{\Theta },[\mathcal{V}_{(\Theta ,\mathcal{F}_{\Theta })}]).$
From our above discussion it follows that $\mathcal{G}_{(\Theta ,\mathcal{F}%
_{\Theta })}=\emptyset $, whenever $\mathcal{F}_{\Lambda }=\{\emptyset
,\Lambda \},$ or $\mathcal{F}_{\Theta }=\{\emptyset ,\Theta \}.$ The latter
relation implies that on\textit{\ a trivial information state space all
generalized observables are trivial}.

For our further needs in section 3.5, we specify a special type of a
generalized observable.

\begin{definition}
\textit{We say that a generalized observable, with an outcome space\ }$%
(\Lambda ,\mathcal{F}_{\Lambda })$ \textit{and on} $(\Theta ,\mathcal{F}%
_{\Theta }),$ \textit{represents an} \textbf{observable} \textit{and, for
specificity, we denote an observable by} $\mathrm{E},$\textit{\ if, for any} 
$B_{0}\in \mathcal{F}_{\Lambda },$ $\mathrm{E}(B_{0})\neq 0,$ \textit{there
exists an element} $\theta _{0}\in \Theta $ \textit{such that}: 
\begin{eqnarray*}
(\mathrm{E}(B))(\theta _{0}) &=&1,\text{ \ \ \ }B\supseteq B_{0},\text{ \ \
\ }\forall B\in \mathcal{F}_{\Lambda }, \\
(\mathrm{E}(B))(\theta _{0}) &=&0,\text{ \ \ \ }B\cap B_{0}=\emptyset ,\text{
\ }\forall B\in \mathcal{F}_{\Lambda }.
\end{eqnarray*}
\end{definition}

Let $\Pi _{1}$ and $\Pi _{2}$ be generalized observables on $(\Theta ,%
\mathcal{F}_{\Theta }),$ with outcome spaces $(\Lambda _{1},\mathcal{F}%
_{\Lambda _{1}})$ and $(\Lambda _{2},\mathcal{F}_{\Lambda _{2}}),$
respectively.

To any generalized observables $\Pi _{1}$ and $\Pi _{2}$, there exists the
unique generalized observable $\Pi _{1}\times \Pi _{2}$ on $(\Theta ,%
\mathcal{F}_{\Theta }),$ with the product outcome space $(\Lambda _{1}\times
\Lambda _{2},\mathcal{F}_{\Lambda _{1}}\otimes \mathcal{F}_{\Lambda _{2}}),$
such that 
\begin{equation*}
((\Pi _{1}\times \Pi _{2})(B_{1}\times B_{2}))(\theta )=(\Pi
_{1}(B_{1}))(\theta )(\Pi _{2}(B_{2}))(\theta ),
\end{equation*}
for all $\theta \in \Theta ,$ $B_{1}\in \mathcal{F}_{\Lambda _{1}},$ $%
B_{2}\in \mathcal{F}_{\Lambda _{2}}.$

We call $\Pi _{1}\times \Pi _{2}$ the \textit{product generalized observable}
of $\Pi _{1}$ and $\Pi _{2}.$

We say that a generalized observable $\Pi $ on $(\Theta ,\mathcal{F}_{\Theta
}),$ with the outcome product space $(\Lambda _{1}\times \Lambda _{2},%
\mathcal{F}_{\Lambda _{1}}\otimes \mathcal{F}_{\Lambda _{2}}),$ is a \textit{%
joint generalized} \textit{observable} of $\Pi _{1}$ and $\Pi _{2}$ if the
latter are the marginal measures of $\Pi $, that is: 
\begin{eqnarray*}
\Pi _{1}(B_{1}) &=&\Pi (B_{1}\times \Lambda _{2}),\text{ \ \ \ }\forall
B_{1}\in \mathcal{F}_{\Lambda _{1}}, \\
\Pi _{2}(B_{2}) &=&\Pi (\Lambda _{1}\times B_{2}),\text{ \ \ \ }\forall
B_{2}\in \mathcal{F}_{\Lambda _{2}}.
\end{eqnarray*}
In particular, the product generalized observable $\Pi _{1}\times \Pi _{2}$
represents a joint generalized observable of $\Pi _{1}$ and $\Pi _{2}.$

However, if generalized observables $\Pi _{1}$ and $\Pi _{2}$ on $(\Theta ,%
\mathcal{F}_{\Theta })$ belong to some definite class then the product
generalized observable may not belong to this class, and, hence, in this
class there may not exist a joint generalized observable of $\Pi _{1}$ and $%
\Pi _{2}$.

Let, for example, a measurable space ($\Theta ,\mathcal{F}_{\Theta })$ has
the property (\ref{14}) and both $\Pi _{1}$ and $\Pi _{2}$ belong to the
class $\mathcal{G}_{(\Theta ,\mathcal{F}_{\Theta })}^{(lin)}$ of non-trivial
generalized observables, represented by measures with values that are
non-negative continuous \textit{linear} functionals on $\mathrm{V}$. Then
the product generalized observable $\Pi _{1}\times \Pi _{2}$ does not belong
to the class $\mathcal{G}_{(\Theta ,\mathcal{F}_{\Theta })}^{(lin)}.$

\subsubsection{Convolution of generalized observables}

Consider measurable spaces $(\Theta ,\mathcal{F}_{\Theta })$ and $(\Theta
^{\prime },\mathcal{F}_{\Theta ^{\prime }}),$ and let $\Pi $ be a
generalized observable on $(\Theta ,\mathcal{F}_{\Theta }),$ with an outcome
space $(\Lambda ,\mathcal{F}_{\Lambda }),$ and \textrm{S} be a generalized
observable on $(\Theta ^{\prime },\mathcal{F}_{\Theta ^{\prime }})$, with an
outcome space $(\Theta ,\mathcal{F}_{\Theta })$.

\begin{definition}
\textit{We call a generalized observable} \textit{on} $(\Theta ^{\prime },%
\mathcal{F}_{\Theta ^{\prime }}),$\textit{\ defined by the relation} 
\begin{equation}
(\Pi (B)\ast \mathrm{S})(\theta ^{\prime }):=\int_{\Theta }(\Pi (B))(\theta
)(\mathrm{S}(d\theta ))(\theta ^{\prime }),\text{ \ \ \ \ }\forall B\in 
\mathcal{F}_{\Lambda },\text{ \ \ }\forall \theta ^{\prime }\in \Theta
^{\prime },  \label{16}
\end{equation}
the\textit{\ \textbf{convolution} of a generalized observable\ }$\Pi ,$ 
\textit{with an outcome space} $(\Lambda ,\mathcal{F}_{\Lambda })$ and 
\textit{on} $(\Theta ,\mathcal{F}_{\Theta }),$ \textit{and\ a generalized
observable} $\mathrm{S},$ \textit{with the outcome space }$(\Theta ,\mathcal{%
F}_{\Theta })$\textit{\ and }on $(\Theta ^{\prime },\mathcal{F}_{\Theta
^{\prime }})$
\end{definition}

If $\phi :\Theta ^{\prime }\rightarrow \Theta $ is an $\mathcal{F}_{\Theta
^{\prime }}/\mathcal{F}_{\Theta }$-measurable function and 
\begin{equation*}
(\widetilde{\mathrm{S}}(F))(\theta ^{\prime })=\chi _{\phi ^{-1}(F)}(\theta
^{\prime }),\ \ \ \forall \theta ^{\prime }\in \Theta ^{\prime },\ \ \forall
F\in \mathcal{F}_{\Theta },
\end{equation*}
then 
\begin{equation*}
(\Pi (B)\ast \widetilde{\mathrm{S}})(\theta ^{\prime })=(\Pi (B))(\phi
(\theta ^{\prime }))=(\Pi (B)\circ \phi )(\theta ^{\prime }),
\end{equation*}
for all $B\in \mathcal{F}_{\Lambda },$ $\theta ^{\prime }\in \Theta ^{\prime
}.$

\begin{definition}
\textit{We say that a generalized observable on} $(\Theta ^{\prime },%
\mathcal{F}_{\Theta ^{\prime }}),$ \textit{defined by} the relation 
\begin{equation}
\Pi _{\phi ^{-1}}(B):=\Pi (B)\circ \phi ,\text{ \ \ \ \ }\forall B\in 
\mathcal{F}_{\Lambda },  \label{17}
\end{equation}
is the $\phi $-\textbf{preimage} \textit{on} $(\Theta ^{\prime },\mathcal{F}%
_{\Theta ^{\prime }})$ \textit{of a generalized observable} $\Pi $ \textit{on%
} $(\Theta ,\mathcal{F}_{\Theta }).$
\end{definition}

For a trivial generalized observable $\Pi $ on $(\Theta ,\mathcal{F}_{\Theta
}),$ its preimage $\Pi _{\phi ^{-1}}$ on $(\Theta ^{\prime },\mathcal{F}%
_{\Theta ^{\prime }})$ is also trivial. Let a generalized observable $\Pi $
be itself an $f$-preimage of some generalized observable $\widetilde{\Pi }$
on $(\widetilde{\Theta },\mathcal{F}_{\widetilde{\Theta }}),$ that is, $\Pi =%
\widetilde{\Pi }_{f^{-1}},$ for an $\mathcal{F}_{\Theta }/\mathcal{F}_{%
\widetilde{\Theta }}$-measurable function $f:\Theta \rightarrow \widetilde{%
\Theta },$ then 
\begin{equation*}
\Pi _{\phi ^{-1}}=\widetilde{\Pi }_{g^{-1}},\text{ \ \ \ \ \ }g=f\circ \phi ,
\end{equation*}
and $\Pi _{\phi ^{-1}}$ is the $g$-preimage on $(\Theta ^{\prime },\mathcal{F%
}_{\Theta ^{\prime }})$ of the generalized observable $\widetilde{\Pi }$ on $%
(\widetilde{\Theta },\mathcal{F}_{\widetilde{\Theta }}).$

\begin{proposition}
\textit{If }$\Pi _{\phi ^{-1}}\ $\textit{on} $(\Theta ^{\prime },\mathcal{F}%
_{\Theta ^{\prime }})$ \textit{is an observable then }$\Pi $ \textit{on }$%
(\Theta ,\mathcal{F}_{\Theta })$ \textit{is also an observable}.
\end{proposition}

\begin{proposition}
\textit{Let} \textit{an }$\mathcal{F}_{\Theta ^{\prime }}/\mathcal{F}%
_{\Theta }$-\textit{measurable function} $\phi :\Theta ^{\prime }\rightarrow
\Theta $ \textit{be surjective}$.$ \textit{Then}:\smallskip \newline
\textbf{(i)} $\Pi $ \textit{on }$(\Theta ,\mathcal{F}_{\Theta })$ \textit{is
trivial iff }$\Pi _{\phi ^{-1}}$ on $(\Theta ^{\prime },\mathcal{F}_{\Theta
^{\prime }})$ \textit{is trivial};\newline
(\textbf{ii}$)$ $\Pi $\textit{\ is an observable on} $(\Theta ,\mathcal{F}%
_{\Theta })$ \textit{iff }$\Pi _{\phi ^{-1}}$ \textit{is an observable on} $%
(\Theta ^{\prime },\mathcal{F}_{\Theta ^{\prime }})$.
\end{proposition}

\subsubsection{Functional subordination}

Consider a generalized observable $\Pi :$ $\mathcal{F}_{\Lambda }\rightarrow 
\frak{B}_{+}\frak{(}\Theta ),$ $\Pi (\Lambda )=\mathcal{I}_{_{\frak{B(}%
\Theta )}},$ on $(\Theta ,\mathcal{F}_{\Theta })$ with an outcome space $%
(\Lambda ,\mathcal{F}_{\Lambda }).$

Let $(\widetilde{\Lambda },\mathcal{F}_{\widetilde{\Lambda }})$ be a
measurable space and $\varphi :\Lambda \rightarrow \widetilde{\Lambda }$ be
an $\mathcal{F}_{\Lambda }/\mathcal{F}_{\widetilde{\Lambda }}$-measurable
function.

\begin{definition}
\textit{We say that on }$(\Theta ,\mathcal{F}_{\Theta })$\textit{\ a
generalized observable }$\widetilde{\Pi }$\textit{, with an outcome space} $(%
\widetilde{\Lambda },\mathcal{F}_{\widetilde{\Lambda }}),$ \textit{is }$%
\varphi $\textit{-functionally subordinated to a generalized observable }$%
\Pi ,$\textit{\ with an outcome space }$(\Lambda ,\mathcal{F}_{\Lambda }),$ 
\textit{and denote this by} $\widetilde{\Pi }=\Pi \circ \varphi ^{-1},$%
\textit{\ if} 
\begin{equation}
(\Pi \circ \varphi ^{-1})(\widetilde{B}):=\Pi (\varphi ^{-1}(\widetilde{B})),
\label{18}
\end{equation}
\textit{for} all $\widetilde{B}\in \mathcal{F}_{\widetilde{\Lambda }}.$
\end{definition}

Consider an $\phi $-preimage $(\Pi \circ \varphi ^{-1})_{\phi ^{-1}}$ on $%
(\Theta ^{\prime },\mathcal{F}_{\Theta ^{\prime }})$ of a generalized
observable $\Pi \circ \varphi ^{-1}$ on $(\Theta ,\mathcal{F}_{\Theta }).$
Due to (\ref{17}) and (\ref{18}), we have 
\begin{eqnarray*}
(\Pi \circ \varphi ^{-1})_{\phi ^{-1}}(\widetilde{B}) &=&(\Pi \circ \varphi
^{-1})(\widetilde{B})\circ \phi =\Pi (\varphi ^{-1}(\widetilde{B}))\circ \phi
\\
&=&\Pi _{\phi ^{-1}}(\varphi ^{-1}(\widetilde{B}))=(\Pi _{\phi ^{-1}}\circ
\varphi ^{-1})(\widetilde{B}),
\end{eqnarray*}
for all $\widetilde{B}\in \mathcal{F}_{\widetilde{\Lambda }}$\ 

\begin{conclusion}
An $\phi $-preimage subordination preserves a $\varphi $-functional
subordination.
\end{conclusion}

\subsection{Outcome probability laws}

In view of the notions, introduced in sections 3.2-3.4, and theorem 1, let
us come back to the general settings of section 3.1.

Consider an experiment $\mathcal{E}$ with outcomes in $(\Omega ,\mathcal{F}%
_{\Omega })$ upon a system $\mathcal{S}$ of any type. Suppose that, before
this experiment, a system $\mathcal{S}$ is represented by an information
state space $(\Theta ,\mathcal{F}_{\Theta },[\mathcal{V}_{(\Theta ,\mathcal{F%
}_{\Theta })}]).$

Let $\mu _{\mathcal{E}}(\cdot ;\cdot )$ be the mapping, specified by (\ref{8}%
), (\ref{9}), (\ref{10}). For each $\pi \subset \mathcal{V}_{(\Theta ,%
\mathcal{F}_{\Theta })},$ the mapping $\mu _{\mathcal{E}}(\cdot ;\pi )$
defines the outcome probability law of the experimental situation $\mathcal{E%
}+\mathcal{S}(\pi ).$ The mapping 
\begin{equation*}
\widetilde{\mu }_{\mathcal{E}}:\mathcal{F}_{\Omega }\rightarrow \frak{B}_{+}%
\frak{(}[\mathcal{V}_{(\Theta ,\mathcal{F}_{\Theta })}]),
\end{equation*}
associated to $\mu _{\mathcal{E}}(\cdot ;\cdot )$ by the relation 
\begin{equation*}
(\widetilde{\mu }_{\mathcal{E}}(\cdot ))([\pi ]):=\mu _{\mathcal{E}}(\cdot
,\pi ),\text{ \ \ }\forall \pi \subset \mathcal{V}_{(\Theta ,\mathcal{F}%
_{\Theta })},
\end{equation*}
satisfies the conditions of theorem 1. Hence, we have the following
proposition.

\begin{theorem}[Representation theorem]
\textit{To} \textit{any} \textit{experiment} $\mathcal{E},$ \textit{with an
outcome space} $(\Omega ,\mathcal{F}_{\Omega }),$ \textit{upon a system }$%
\mathcal{S}$ \textit{represented initially by} \textit{an information state
space }$(\Theta ,\mathcal{F}_{\Theta },[\mathcal{V}_{(\Theta ,\mathcal{F}%
_{\Theta })}]),$ \textit{there corresponds a unique} \textit{generalized} 
\textit{observable }$\Pi $ \textit{on} $(\Theta ,\mathcal{F}_{\Theta })$%
\textit{\ such that\ the outcome probability law of each experimental
situation }$\mathcal{E}+\mathcal{S}(\pi ),$ $\pi \in \mathcal{V}_{(\Theta ,%
\mathcal{F}_{\Theta })},$ \textit{is given by} 
\begin{equation}
\mu _{\mathcal{E}}(B;\pi )=\int_{\Theta }(\Pi (B))(\theta )\pi (d\theta ),
\label{19}
\end{equation}
\textit{for all} $B\in \mathcal{F}_{\Omega }.$
\end{theorem}

\begin{remark}
Clearly, the converse statement is not true and, to a generalized observable%
\textit{\ }$\Pi $\textit{\ }on $(\Theta ,\mathcal{F}_{\Theta })$ there may,
in general, correspond a variety of experiments upon $\mathcal{S},$\textit{\ 
}with the same probability distribution of outcomes. We further denote by $%
[\Pi ]$ the equivalence class of all experiments upon $\mathcal{S}$,
represented on $(\Theta ,\mathcal{F}_{\Theta })$ by the same non-trivial
generalized observable $\Pi .$
\end{remark}

In the terminology of section 3.1, the relation (\ref{19}) implies that, for
each experimental situation $\mathcal{E}+\mathcal{S}(\pi ),$ $\pi \in 
\mathcal{V}_{(\Theta ,\mathcal{F}_{\Theta })},$ the probability space 
\begin{equation*}
(\Omega ,\mathcal{F}_{\Omega },\mu _{\mathcal{E}}(\cdot ;\pi ))
\end{equation*}
is \textit{induced} by an initial information state $(\Theta ,\mathcal{F}%
_{\Theta },[\pi ])$.

\subsubsection{''No knowledge''}

Notice that if:

(i) an initial information state space is trivial, then on this space there
exist only trivial generalized observables and, in this case, any experiment
upon $\mathcal{S}$ is represented on this space by a trivial generalized
observable;

(ii) in an outcome space $(\Omega ,\mathcal{F}_{\Omega })$, the $\sigma $%
-algebra $\mathcal{F}_{\Omega }$ is trivial, then there exists only one
normalized positive scalar measure $\pi _{0}$ in $\mathcal{V}_{(\Omega ,%
\mathcal{F}_{\Omega })}.$ In this case, in (\ref{19}), $\mu _{\mathcal{E}%
}(\cdot ;\pi )=\pi _{0}(\cdot ),$\ $\forall \pi \in \mathcal{V}_{(\Theta ,%
\mathcal{F}_{\Theta })}$ and the generalized observable on $(\Theta ,%
\mathcal{F}_{\Theta }),$ corresponding to this experiment, is given by $\Pi
=\pi _{0}\mathcal{I}_{_{\frak{B(}\Theta )}},$ and is also trivial.

(iii) $(\Theta ,\mathcal{F}_{\Theta },[\mathcal{V}_{(\Theta ,\mathcal{F}%
_{\Theta })}])$ is a non-trivial initial information state space of $%
\mathcal{S}$ and 
\begin{equation*}
\mu _{\mathcal{E}}(\cdot ;\pi )=\mu _{\mathcal{E}}(\cdot ),\text{ \ \ }%
\forall \pi \in \mathcal{V}_{(\Theta ,\mathcal{F}_{\Theta })},
\end{equation*}
then, in (\ref{19}), the corresponding generalized observable is trivial.

Due to our discussion in section 3.1, in all above cases the initial
information on $\mathcal{S}$, represented by $(\Theta ,\mathcal{F}_{\Theta
},[\mathcal{V}_{(\Theta ,\mathcal{F}_{\Theta })}]),$ provides ''\textit{no} 
\textit{knowledge}'' for the description of an experiment $\mathcal{E}$.

\begin{proposition}[''No knowledge'']
\textit{An information state space} $(\Theta ,\mathcal{F}_{\Theta },[%
\mathcal{V}_{(\Theta ,\mathcal{F}_{\Theta })}])$ \textit{of} $\mathcal{S}$ 
\textit{provides }''\textit{no knowledge}'' \textit{on} \textit{an
experiment }$\mathcal{E}$ \textit{iff the generalized observable }$\Pi ,$ 
\textit{representing on} $(\Theta ,\mathcal{F}_{\Theta })$ \textit{this
experiment,} \textit{is trivial.}
\end{proposition}

From this proposition it follows that a trivial information state space
provides ''no knowledge'' on experiments upon $\mathcal{S}.$ A non-trivial
information state space $(\Theta ,\mathcal{F}_{\Theta },[\mathcal{V}%
_{(\Theta ,\mathcal{F}_{\Theta })}])$ \textit{provides the knowledge only on
the class of experiments, represented on }$(\Theta ,\mathcal{F}_{\Theta })$%
\textit{\ by non-trivial generalized observables, that is, those in }$%
\mathcal{G}_{(\Theta ,\mathcal{F}_{\Theta })}.$

If an information state space of $\mathcal{S}$ is obvious from the context,
we further refer to $\Pi $ as a generalized observable of $\mathcal{S}$.

\subsubsection{Informational equivalence}

Consider the situation where an initial information state $(\Theta ,\mathcal{%
F}_{\Theta },[\pi ])$ of $\mathcal{S}$ is \textit{induced}, due to (\ref{11}%
), by an information state $(\Theta ^{\prime },\mathcal{F}_{\Theta ^{\prime
}},[\pi ^{\prime }]).$ In terms of generalized observables, the relation (%
\ref{11}) implies that there exists a generalized observable $\mathrm{S,}$
with the outcome space $(\Theta ,\mathcal{F}_{\Theta })$ and on $(\Theta
^{\prime },\mathcal{F}_{\Theta ^{\prime }})$, such that 
\begin{equation}
\pi (F)=\int_{\Theta ^{\prime }}(\mathrm{S}(F))(\theta ^{\prime })\pi
^{\prime }(d\theta ^{\prime }),\text{ \ \ }\forall F\in \mathcal{F}_{\Theta
}.  \label{20}
\end{equation}

Let $\mathcal{E}$ be an experiment upon $\mathcal{S},$ represented on $%
(\Theta ,\mathcal{F}_{\Theta })$ by a non-trivial generalized observable $%
\Pi .$ \ Consider the probability distribution of an experimental situation $%
\mathcal{E}+\mathcal{S}(\pi ),$ with $\pi $ being defined by (\ref{20}).
From (\ref{20}) and (\ref{19}) it follows that 
\begin{equation}
\mu _{\mathcal{E}}(B;\pi )=\int_{\Theta }(\Pi (B))(\theta )\pi (d\theta
)=\int_{\Theta ^{\prime }}(\Pi ^{\prime }(B))(\theta ^{\prime })\pi ^{\prime
}(d\theta ^{\prime }),\text{ \ \ \ \ \ \ \ \ }\forall B\in \mathcal{F}%
_{\Omega },  \label{21}
\end{equation}
where $\Pi ^{\prime }$ is a generalized observable 
\begin{equation}
(\Pi ^{\prime }(B))(\theta ^{\prime })=(\Pi (B)\ast \mathrm{S})(\theta
^{\prime }),\text{ \ \ \ \ }\forall B\in \mathcal{F}_{\Omega },  \label{22}
\end{equation}
\textit{generated }on $(\Theta ^{\prime },\mathcal{F}_{\Theta ^{\prime }})$
by generalized observables\textit{\ }$\Pi $ on $(\Theta ^{\prime },\mathcal{F%
}_{\Theta ^{\prime }})$ and $\mathrm{S}$ on $(\Theta ^{\prime },\mathcal{F}%
_{\Theta ^{\prime }})$ and representing the \textit{convolution} of these
generalized observables.

The relation (\ref{21}) shows that if an information state $(\Theta ,%
\mathcal{F}_{\Theta },[\pi ])$ is induced by an information state $(\Theta
^{\prime },\mathcal{F}_{\Theta ^{\prime }},[\pi ^{\prime }])$ then, to an
experimental situation $\mathcal{E}_{\Pi }\mathcal{+S}(\pi ),$ there
corresponds the experimental situation $\mathcal{E}_{\Pi ^{\prime }}^{\prime
}\mathcal{+S}(\pi ^{\prime }),$ represented by the generated generalized
observable $\Pi ^{\prime }$ on $(\Theta ^{\prime },\mathcal{F}_{\Theta
^{\prime }})$, defined by (\ref{22}). In view of (\ref{21}), we denote by 
\begin{equation*}
\mathcal{E}_{\Pi }\mathcal{+S}(\pi )\thicksim \mathcal{E}_{\Pi ^{\prime
}}^{\prime }\mathcal{+S}(\pi ^{\prime })
\end{equation*}
the informational equivalence of these experimental situations.

\subsubsection{Deterministic set-up}

Let a system $\mathcal{S}$ be initially represented by an information state
space $(\Theta ,\mathcal{F}_{\Theta },[\mathcal{V}_{(\Theta ,\mathcal{F}%
_{\Theta })}]).$

To separate the cases where an experiment $\mathcal{E}$ upon a system $%
\mathcal{S}$ may have a probabilistic setup (see point (ii) in section 3.1),
consider the situation where all atom subsets $\{\theta \}$ of $\Theta $
belong to $\mathcal{F}_{\Theta }$ and, for, at least, one pure information
state $(\Theta ,\mathcal{F}_{\Theta },[\delta _{\theta _{0}}])$ of $\mathcal{%
S},$ the predictions are formulated in the language of ''yes - no''
statements. Since a set-up of an experiment does not depend on an initial
information state of $\mathcal{S},$ the considered experiment has a
deterministic set-up\footnote{%
See also the discussion in [9,11].}.

Due to definition 7, this experiment is represented on $(\Theta ,\mathcal{F}%
_{\Theta },[\mathcal{V}_{(\Theta ,\mathcal{F}_{\Theta })}])$ by an \textit{%
observable}\emph{\ }$\mathrm{E}$ and from (\ref{19}) it follows that the
outcome probability law of each experimental situation $\mathcal{E}+\mathcal{%
S}(\pi ),$ $\pi \in \mathcal{V}_{(\Theta ,\mathcal{F}_{\Theta })}$, is given
by 
\begin{equation*}
\mu (B;\pi )=\int_{\Theta }(\mathrm{E}(B))(\theta )\pi (d\theta ),\text{ \ \ 
}\forall B\in \mathcal{F}_{\Lambda }.
\end{equation*}

Thus, \textit{on a system information state space an observable represents
an experiment with a deterministic set-up}.

In general, a non-trivial observable represents an experiment $\mathcal{E}$
upon $\mathcal{S}$ as a whole and can not be referred to some property of $%
\mathcal{S},$ existing before this experiment. However, the latter is true
for the special type of observables, which we introduce in section 4.3.1\
and call \textit{beables}.

\begin{example}
Let $\varphi :\Theta \rightarrow \Omega $ be an $\mathcal{F}_{\Theta }/%
\mathcal{F}_{\Omega }$-measurable function. Consider on $(\Theta ,\mathcal{F}%
_{\Theta })$ an observable 
\begin{equation}
(\mathrm{E}(B))(\theta )=\chi _{\varphi ^{-1}(B)}(\theta ),\text{ \ \ \ }%
\forall \theta \in \Theta ,\text{ \ \ }\forall B\in \mathcal{F}_{\Omega },
\label{23}
\end{equation}
with an outcome space $(\Omega ,\mathcal{F}_{\Omega })$. For each
experimental situation $\mathcal{E}+\mathcal{S}(\pi ),$ $\pi \in \mathcal{V}%
_{(\Theta ,\mathcal{F}_{\Theta })},$ the probability distribution of
outcomes in $(\Omega ,\mathcal{F}_{\Omega })$ is given by 
\begin{equation}
\mu (B;\pi )=\pi (\varphi ^{-1}(B))=(\pi \circ \varphi ^{-1})(B),\text{ \ \ }%
\forall B\in \mathcal{F}_{\Omega },  \label{24}
\end{equation}
and, hence, the probability space $(\Omega ,\mathcal{F}_{\Omega },\mu (\cdot
;\pi ))$ is the $\varphi $-image of the initial information state $(\Theta ,%
\mathcal{F}_{\Theta },[\pi ]),$ that is: 
\begin{equation*}
(\Omega ,\mathcal{F}_{\Omega },\mu (\cdot ;\pi ))=\varphi \lbrack (\Theta ,%
\mathcal{F}_{\Theta },[\pi ])].
\end{equation*}
In the frame of conventional probability theory, the probability
distribution (\ref{24}) is called an\ image probability law while an $%
\mathcal{F}_{\Theta }$/$\mathcal{F}_{\Omega }$-measurable mapping $\varphi $
is called a random variable on\ $(\Omega ,\mathcal{F}_{\Omega }).$\ It is
assumed that the image law predicts the probability distribution of outcomes
under a non-perturbing (by device) experiment, with a deterministic set-up,
representing a classical ''errorless''\textit{\ }measurement\footnote{%
See, for example, the discussion in [11].} of the property $\varphi $ of $%
\mathcal{S},$ existing objectively before this measurement. \newline
Let $\sigma $-algebras $\mathcal{F}_{\Theta }$ and $\mathcal{F}_{\Omega }$
contain all atom subsets. In this case, under an experiment, described by (%
\ref{23}), to each initial pure information state $(\Theta ,\mathcal{F}%
_{\Theta },[\delta _{\theta _{0}}]),$ $\theta _{0}\in \Theta ,$ the outcome $%
\omega _{0}=\varphi (\theta _{0})$ is predicted with certainty. However,
since, immediately after this experiment, the description of the system in
terms of information states is not specified, one can not claim\textit{\ }%
that the observable\textit{\ }(\ref{23}) represents a non-perturbing
experiment on the measurement of a property $\varphi $\textit{\ }of\textit{\ 
}$\mathcal{S}.$ To specify when this is really the case, we introduce in
section 4.3 the notion of a non-perturbing experiment upon a system $%
\mathcal{S}.$
\end{example}

\section{Non-destructive experiments}

According to section 3.5.1,\ an information state space\textit{\ }$(\Theta ,%
\mathcal{F}_{\Theta },[\mathcal{V}_{(\Theta ,\mathcal{F}_{\Theta })}])$
provides the knowledge on the description of only such class of\textit{\ }%
experiments upon\textit{\ $\mathcal{S}$ }which are represented on $(\Theta ,%
\mathcal{F}_{\Theta })$ by non-trivial generalized observables, that is, the
generalized observables in\textit{\ }$\mathcal{G}_{(\Theta ,\mathcal{F}%
_{\Theta })}$.

In this section we introduce, in the most general settings, the concepts of
a complete information description and a complete statistical description of
a non-destructive experiment. We define the notion of a non-perturbing
experiment and discuss the phenomenon of ''reduction'' of an information
state.

\subsection{Extended generalized observables}

Let the initial information on a system $\mathcal{S}$ be described by $%
(\Theta _{in},\mathcal{F}_{\Theta _{in}},[\mathcal{V}_{(\Theta _{in},%
\mathcal{F}_{\Theta _{in}})}]),$ with a non-empty set $\mathcal{G}_{(\Theta
_{in},\mathcal{F}_{\Theta _{in}})}.$

Consider the description of an experiment upon $\mathcal{S}$ such that,
immediately after this experiment, the system $\mathcal{S}$ exists. We call
such experiments \textit{non-destructive}.

Let\ a non-destructive experiment $\mathcal{E}$, with outcomes in $(\Omega ,%
\mathcal{F}_{\Omega }),$ upon $\mathcal{S}$ be represented on $(\Theta _{in},%
\mathcal{F}_{\Theta _{in}})$ by a non-trivial generalized observable.
Suppose that, immediately after this non-destructive experiment, a system $%
\mathcal{S}$ is characterized in terms of an output information state space 
\begin{equation*}
(\Theta _{out},\mathcal{F}_{\Theta _{out}},[\mathcal{V}_{(\Theta _{out},%
\mathcal{F}_{\Theta _{out}})}]),
\end{equation*}
with a non-empty set $\mathcal{G}_{(\Theta _{out},\mathcal{F}_{\Theta
_{out}})},$ and, in general, different from the initial information state
space of $\mathcal{S}.$

Each trial of an experimental situation $\mathcal{E}+\mathcal{S}(\pi _{in}),$
$\pi _{in}\in \mathcal{V}_{(\Theta _{in},\mathcal{F}_{\Theta _{in}})},$
results in a outcome $\omega \in \Omega $ and a posterior system $\mathcal{S}%
,$ represented by a posterior $\mathcal{S}$-outcome $\theta _{out}$ in $%
(\Theta _{out},\mathcal{F}_{\Theta _{out}}).$ Under a non-destructive
experiment $\mathcal{E}$ upon $\mathcal{S},$ the pair $(\omega ,\theta
_{out})$ represents a compound outcome in the extended outcome space 
\begin{equation*}
(\Omega \times \Theta _{out},\mathcal{F}_{\Omega }\otimes \mathcal{F}%
_{\Theta _{out}}).
\end{equation*}

Denote by 
\begin{equation}
\nu (\cdot ;\pi _{in}):\text{ }\mathcal{F}_{\Omega }\otimes \mathcal{F}%
_{\Theta _{out}}\rightarrow \lbrack 0,1]  \label{25}
\end{equation}
the probability distribution of compound outcomes $(\omega ,\theta _{out})$
in the extended outcome space\textit{\ }$(\Omega \times \Theta _{out},%
\mathcal{F}_{\Omega }\otimes \mathcal{F}_{\Theta _{out}})$\textit{. }Notice
that $\nu (\cdot ;\pi _{in}^{\prime })=\nu (\cdot ;\pi _{in}),$ $\forall \pi
_{in}^{\prime }\in \lbrack \pi _{in}].$

It is also natural to assume that, for the probability distribution $\nu
(\cdot ;\cdot )$, the statistical axiom,\textit{\ }section\ 3.1\textit{, }is
valid

Consider\textit{\ }the mapping\textit{\ }$(\widetilde{\nu }(\cdot ))(\cdot
), $ associated to the mapping (\ref{25})\textit{\ }by the relation 
\begin{equation*}
(\widetilde{\nu }(\cdot ))([\pi _{in}]):=\nu (\cdot ;\pi _{in}),\ \text{\ \
\ \ \ }\forall \pi _{in}\subset \mathcal{V}_{(\Theta _{in},\mathcal{F}%
_{\Theta _{in}})}.
\end{equation*}
Then 
\begin{equation*}
(\widetilde{\nu }(\cdot ))(\cdot ):(\mathcal{F}_{\Omega }\otimes \mathcal{F}%
_{\Theta _{out}})\times \lbrack \mathcal{V}_{(\Theta _{in},\mathcal{F}%
_{\Theta _{in}})}]\rightarrow \lbrack 0,1]
\end{equation*}
satisfies the conditions of theorem 1 and, hence, to $\widetilde{\nu },$
there corresponds the uniquely defined generalized observable 
\begin{equation*}
\Upsilon :\mathcal{F}_{\Omega }\otimes \mathcal{F}_{\Theta
_{out}}\rightarrow \frak{B}_{+}(\Theta _{in}),
\end{equation*}
with the outcome space $(\Omega \times \Theta _{out},\mathcal{F}_{\Omega
}\otimes \mathcal{F}_{\Theta _{out}})$ and on $(\Theta _{in},\mathcal{F}%
_{\Theta _{in}}),$ such that 
\begin{equation*}
\nu (B;\pi _{in})=\int_{\Theta _{in}}(\Upsilon (B))(\theta _{in})\pi
_{in}(d\theta _{in}),
\end{equation*}
for all $\pi _{in}\in \mathcal{V}_{(\Theta _{in},\mathcal{F}_{\Theta
_{in}})},$ $B\in \mathcal{F}_{\Omega }.$

The marginal generalized observables 
\begin{eqnarray*}
\mathrm{M}_{\Upsilon }(B) &:&=\Upsilon (B\times \Theta _{out}),\text{ \ \ }%
\forall B\in \mathcal{F}_{\Omega }, \\
\mathrm{S}_{\Upsilon }(F_{out}) &:&=\Upsilon (\Omega \times F_{out}),\text{
\ }\ \forall F_{out}\in \mathcal{F}_{\Theta _{out}},
\end{eqnarray*}
define, for each $\mathcal{E}+\mathcal{S}(\pi _{in}),$ $\pi _{in}\in 
\mathcal{V}_{(\Theta _{in},\mathcal{F}_{\Theta _{in}})},$ the probability
distribution 
\begin{equation}
\mu _{\Upsilon }(B;\pi _{in}):=\mu _{\mathrm{M}_{\Upsilon }}(B;\pi
_{in})=\int_{\Theta _{in}}(\mathrm{M}_{\Upsilon }(B))(\theta _{in})\pi
_{in}(d\theta _{in}),\text{ \ }\forall B\in \mathcal{F}_{\Omega },
\label{26}
\end{equation}
of outcomes $\omega $ in $(\Omega ,\mathcal{F}_{\Omega })$ and the
unconditional probability distribution 
\begin{equation}
\tau _{\Upsilon }(F_{out};\pi _{in}):=\tau _{\mathrm{S}_{\Upsilon
}}(F_{out};\pi _{in})=\int_{\Theta _{in}}(\mathrm{S}_{\Upsilon
}(F_{out}))(\theta _{in})\pi _{in}(d\theta _{in}),\text{ \ }\forall
F_{out}\in \mathcal{F}_{\Theta _{out}},  \label{27}
\end{equation}
of posterior $\mathcal{S}$-outcomes $\theta _{out}$ in $(\Theta _{out},%
\mathcal{F}_{\Theta _{out}}),$ in case where outcomes $\omega $ in $(\Omega ,%
\mathcal{F}_{\Omega })$ are ignored completely.

For specificity, we further refer to the generalized observables $\Upsilon ,$
$\mathrm{M}_{\Upsilon },$ $\mathrm{S}_{\Upsilon },$ as an \textit{extended,}
an \textit{outcome} and a \textit{system generalized observable} on\textit{\ 
}$(\Theta _{in},\mathcal{F}_{\Theta _{in}})$, respectively. Since, in our
settings, an outcome generalized observable $\mathrm{M}_{\Upsilon }\in 
\mathcal{G}_{(\Theta _{in},\mathcal{F}_{\Theta _{in}})},$ representing $%
\mathcal{E},$ is non-trivial, the extended generalized observable $\Upsilon $
is also non-trivial, that is, $\Upsilon \in \mathcal{G}_{(\Theta _{in},%
\mathcal{F}_{\Theta _{in}})}$.

\subsection{Complete information description}

For a subset $B\in \mathcal{F}_{\Omega },$ $\mu _{\Upsilon }(B;\pi
_{in})\neq 0,$ the conditional measure 
\begin{equation}
\pi _{\Upsilon }^{out}(F_{out}|B;\pi _{in}):=\frac{\int_{\Theta
_{in}}(\Upsilon (B\times F_{out}))(\theta _{in})\pi _{in}(d\theta _{in})}{%
\mu _{\Upsilon }(B;\pi _{in})}  \label{28}
\end{equation}
defines the probability that, immediately after a single experimental trial $%
\mathcal{E}+\mathcal{S}(\pi _{in}),$ $\pi _{in}\in \mathcal{V}_{(\Theta
_{in},\mathcal{F}_{\Theta _{in}})},$ where only the event that the outcome $%
\omega \in B$ has been recorded, the posterior $\mathcal{S}$-outcome $\theta
_{out}$ belongs to a subset $F_{out}\in \mathcal{F}_{\Theta _{out}}.$

Hence, immediately after this single trial, the information state 
\begin{equation*}
(\Theta _{out},\mathcal{F}_{\Theta _{out}},[\pi _{\Upsilon }^{out}(\cdot
|B;\pi _{in})])
\end{equation*}
represents the \textit{conditional posterior information state}\footnote{%
This terminology is introduced here in view of the similar terminology, used
for the description of experiments upon quantum systems (see, for example,
[17-19]).}\textit{\ }of $\mathcal{S}$ in the output information state space 
\begin{equation*}
(\Theta _{out},\mathcal{F}_{\Theta _{out}},[\mathcal{V}_{(\Theta _{out},%
\mathcal{F}_{\Theta _{out}})}]).
\end{equation*}
The unconditional posterior information state 
\begin{equation*}
(\Theta _{out},\mathcal{F}_{\Theta _{out}},[\pi _{\Upsilon }^{out}(\cdot
|\Omega ;\pi _{in})])=(\Theta _{out},\mathcal{F}_{\Theta _{out}},[\tau
_{\Upsilon }(\cdot ;\pi _{in})])
\end{equation*}
corresponds to the situation where outcomes in $(\Omega ,\mathcal{F}_{\Omega
})$ are ignored completely and only posterior $\mathcal{S}$-outcomes are
considered.

\begin{definition}
\textit{Under the} \textbf{complete information description}\textit{\ of a
non-destructive experiment }$\mathcal{E}$ \textit{upon a system} $\mathcal{S}%
,$ \textit{we mean the\ knowledge, for }each $\pi _{in}\in \mathcal{V}%
_{(\Theta _{in},\mathcal{F}_{\Theta _{in}})},$\textit{\ of the outcome
probability law\ }$\mu _{\mathcal{E}}(\cdot ;\pi _{in})$\textit{\ and the
family} 
\begin{equation*}
\{(\Theta _{out},\mathcal{F}_{\Theta _{out}},[\pi _{\Upsilon }^{out}(\cdot
|B;\pi _{in})]):B\in \mathcal{F}_{\Omega }\}
\end{equation*}
\textit{of all conditional posterior information states of }$\mathcal{S}$%
\textit{\ in }$(\Theta _{out},\mathcal{F}_{\Theta _{out}},[\mathcal{V}%
_{(\Theta _{out},\mathcal{F}_{\Theta _{out}})}])$.
\end{definition}

\noindent Denoting in (\ref{28}) 
\begin{equation*}
((\mathcal{M}_{\Upsilon }^{pr}(B))(\pi _{in}))(F_{out}):=\int_{\Theta
_{in}}(\Upsilon (B\times F_{out}))(\theta _{in})\pi _{in}(d\theta _{in}),
\end{equation*}
we derive 
\begin{equation*}
\pi _{\Upsilon }^{out}(\cdot |B;\pi _{in}))=\frac{((\mathcal{M}_{\Upsilon
}^{pr}(B))(\pi _{in}))(\cdot )}{\mu _{\Upsilon }(B;\pi _{in})}.
\end{equation*}

\begin{definition}
\textit{We call the mapping } 
\begin{equation*}
(\mathcal{M}_{\Upsilon }^{pr}(\cdot ))(\cdot ):\mathcal{F}_{\Omega }\times 
\mathcal{J}_{(\Theta _{in},\mathcal{F}_{\Theta _{in}})}\rightarrow \mathcal{J%
}_{\mathcal{(}\Theta _{out},\mathcal{F}_{\Theta _{out})}},
\end{equation*}
\textit{defined\ to an extended generalized observable }$\Upsilon $ \textit{%
by the relation} 
\begin{equation}
((\mathcal{M}_{\Upsilon }^{pr}(B))(\nu _{in}))(F_{out}):=\int_{\Theta
_{in}}(\Upsilon (B\times F_{out}))(\theta _{in})\nu _{in}(d\theta _{in}),
\label{29}
\end{equation}
for all $B\in \mathcal{F}_{\Omega },$ $F_{out}\in \mathcal{F}_{\Theta
_{out}},$ $\nu _{in}\in \mathcal{J}_{(\Theta _{in},\mathcal{F}_{\Theta
_{in}})},$ \textit{an\textbf{\ information state instrument}.}
\end{definition}

For an information state instrument, the mapping\textit{\ }$\mathcal{M}%
_{\Upsilon }^{pr}(\cdot )$ is a $\sigma $-additive measure on\textit{\ }$%
(\Omega ,\mathcal{F}_{\Omega })$ with values $\mathcal{M}_{\Upsilon
}^{pr}(B),$ $B\in \mathcal{F}_{\Omega },$ that are positive bounded linear
operators\textit{\ }$\mathcal{J}_{(\Theta _{in},\mathcal{F}_{\Theta
_{in}})}\rightarrow \mathcal{J}_{\mathcal{(}\Theta _{out},,\mathcal{F}%
_{\Theta _{out})}}$ and, for each normalized measure $\nu _{in}(\Lambda )=1$%
, the measure\textit{\ }$(\mathcal{M}_{\Upsilon }^{pr}(\Omega ))(\nu _{in})$
on $(\Theta _{out},\mathcal{F}_{\Theta _{out}})$ is also normalized: 
\begin{equation*}
((\mathcal{M}_{\Upsilon }^{pr}(\Omega ))(\nu _{in}))(\Theta _{out})=1.
\end{equation*}
The concept of an information state instrument, defined by (\ref{29}),
coincides with the notion of an instrument, introduced in [6] in case where
the latter is specified for the case of Kolmogorov's model [6, section 4].

From (\ref{26})-(\ref{29}) it follows that, for each experimental situation $%
\mathcal{E}+\mathcal{S}(\pi _{in}),$ $\pi _{in}\in \mathcal{V}_{(\Theta
_{in},\mathcal{F}_{\Theta _{in}})},$ the information state instrument $%
\mathcal{M}_{\Upsilon }^{pr}$ defines by the relations\ 
\begin{eqnarray}
\pi _{\Upsilon }^{out}(F_{out}|B;\pi _{in})) &=&\frac{((\mathcal{M}%
_{\Upsilon }^{pr}(B))(\pi _{in}))(F_{out})}{\mu _{\Upsilon }(B;\pi _{in})},%
\text{ \ \ \ }\mu _{\Upsilon }(B;\pi _{in})\neq 0,  \label{30} \\
\mu _{\Upsilon }(B;\pi _{in}) &=&((\mathcal{M}_{\Upsilon }^{pr}(B))(\pi
_{in}))(\Theta _{out}),  \notag \\
\tau _{\Upsilon }(F_{out};\pi _{in}) &=&((\mathcal{M}_{\Upsilon
}^{pr}(\Omega ))(\pi _{in}))(F_{out}),  \notag
\end{eqnarray}
valid for all $F_{out}\in \mathcal{F}_{\Theta _{out}},B\in \mathcal{F}%
_{\Omega },$ respectively, the probability distribution of outcomes in $%
(\Omega ,\mathcal{F}_{\Omega }),$ the conditional posterior information
states and the unconditional probability distribution of posterior $\mathcal{%
S}$-outcomes in $(\Theta _{out},\mathcal{F}_{\Theta _{out}})$.

\begin{conclusion}
For a non-destructive experiment\textit{\ }$\mathcal{E}$ upon $\mathcal{S},$
represented on an initial information state space of\ $\mathcal{S}$ by a
non-trivial extended generalized observable $\Upsilon $, the complete
information description is given by the notion of an information state
instrument $\mathcal{M}_{\Upsilon }^{pr}.$\medskip
\end{conclusion}

\begin{example}
Consider a non-destructive experiment $\mathcal{E}$ with outcomes in $%
(\Omega ,$ $\mathcal{F}_{\Omega })$ upon $\mathcal{S},$ which is represented
on $(\Theta _{in},\mathcal{F}_{\Theta _{in}})$ by the extended observable 
\begin{equation}
(\mathrm{E}_{(\varphi ,g)}(B\times F_{out}))(\theta _{in})=\chi _{\varphi
^{-1}(B)\cap g^{-1}(F_{out})}(\theta _{in}),  \label{31}
\end{equation}
$\forall B\in \mathcal{F}_{\Omega },$ $F_{out}\in \mathcal{F}_{\Theta
_{out}}.$ Here functions $\varphi :\Theta _{in}\rightarrow \Omega $ and $%
g:\Theta _{in}\rightarrow \Theta _{out}$ are, respectively, $\mathcal{F}%
_{\Theta _{in}}/\mathcal{F}_{\Omega }$ and $\mathcal{F}_{\Theta _{in}}/%
\mathcal{F}_{\Theta _{out}}$ measurable. \newline
To the observable (\ref{31}), the information state instrument, defined, in
general, by (\ref{29}), has the form 
\begin{equation}
((\mathcal{M}_{\mathrm{E}_{(\varphi ,g)}}^{pr}(B))(\pi _{in}))(F_{out})=\pi
_{in}(g^{-1}(F_{out})\cap \varphi ^{-1}(B)),  \label{32}
\end{equation}
for all $\pi _{in}\in \mathcal{V}_{(\Theta _{in},\mathcal{F}_{\Theta
_{in}})} $, $F_{out}\in \mathcal{F}_{\Theta _{out}},$ $B\in \mathcal{F}%
_{\Omega }.$\newline
From (\ref{30}) it follows that, for each experimental situation $\mathcal{E}%
+\mathcal{S}(\pi _{in}),$ $\pi _{in}\in \mathcal{V}_{(\Theta _{in},\mathcal{F%
}_{\Theta _{in}})},$ the probability distribution of outcomes is given by
the image probability distribution 
\begin{equation}
\mu _{\mathrm{E}_{(\varphi ,g)}}(B;\pi _{in})=\pi _{in}(\varphi ^{-1}(B)),%
\text{ \ \ \ }\forall B\in \mathcal{F}_{\Omega },  \label{33}
\end{equation}
while, for each $B\in \mathcal{F}_{\Omega },$ $\pi _{in}(\varphi
^{-1}(B))\neq 0,$ the conditional posterior information state 
\begin{equation*}
(\Theta _{out},\mathcal{F}_{\Theta _{out}},[\pi _{\mathrm{E}_{(\varphi
,g)}}^{out}(\cdot |B;\pi _{in})])
\end{equation*}
is represented by 
\begin{equation}
\pi _{\mathrm{E}_{(\varphi ,g)}}^{out}(F|B;\pi _{in})=\frac{\pi
_{in}(\varphi ^{-1}(B)\cap g^{-1}(F_{out}))}{\pi _{in}(\varphi ^{-1}(B))},%
\text{ \ \ \ }\forall F_{out}\in \mathcal{F}_{\Theta _{out}}.  \label{34}
\end{equation}
\end{example}

\subsubsection{Reduction of an information state}

In the most general settings, for any $B\in \mathcal{F}_{\Omega },$ the
conditional change 
\begin{equation*}
(\Theta _{in},\mathcal{F}_{\Theta _{in}},[\pi _{in}])\mapsto (\Theta _{out},%
\mathcal{F}_{\Theta _{out}},[\pi _{\Upsilon }^{out}(\cdot |B;\pi _{in})]),
\end{equation*}
described, due to (\ref{30}), by the notion of an information state
instrument $\mathcal{M}_{\Upsilon }^{pr},$ represents the phenomenon of
''reduction'' of an initial information state\textit{. }

From our presentation it follows that this phenomenon is inherent, in
general, to any non-destructive experiment and upon a system $\mathcal{S}$
of any type, described in terms of information state spaces.

As we discuss below in section 4.3, in the most general settings, a
reduction of an information state\textit{\ }is induced by:

(i) \textit{the ''renormalization'' of information on a system, conditioned
upon the recorded event under a single experimental trial};

(ii) \textit{the ''dynamical'' change of an information state of a system in
the course of a perturbing experiment}.\smallskip

Since, as we establish in section 6, the probabilistic model of quantum
theory represents a special model of our general framework, the well-known
von Neumann quantum ''state collapse'', postulated in [1], and its further
generalizations (see in [3,13,17,18]), represent particular cases of this
general phenomenon.

\subsection{Non-perturbing experiments}

Consider a non-destructive experiment $\mathcal{E}$ upon a system $\mathcal{S%
}$, represented on $(\Theta _{in},\mathcal{F}_{\Theta _{in}})$ by a
non-trivial extended generalized observable $\Upsilon $.

Introduce the following concept.

\begin{definition}
\textit{We call a non-trivial extended generalized observable }$\Upsilon $%
\textit{\ on }$(\Theta _{in},\mathcal{F}_{\Theta _{in}})$\textit{\
non-perturbing and, for specificity, denote it by} $\Upsilon ^{(np)}$ 
\textit{if there exist:\newline
}(i)\textit{\ a measurable space} $(\Theta ,\mathcal{F}_{\Theta });$\newline
(ii) \textit{an} $\mathcal{F}_{\Theta }/\mathcal{F}_{\Theta _{in}}$-\textit{%
measurable function} $\phi _{in}:\Theta \rightarrow \Theta _{in};$\newline
(iii)\textit{\ an} $\mathcal{F}_{\Theta }/\mathcal{F}_{\Theta _{out}}$-%
\textit{measurable function} $\phi _{out}:\Theta \rightarrow \Theta _{out};$%
\newline
\textit{such that the} $\phi _{in}$-\textit{preimage} $\Upsilon _{\phi
_{in}^{-1}}^{(np)}$ \textit{on} $(\Theta ,\mathcal{F}_{\Theta })$ \textit{%
has the form}: 
\begin{equation}
(\Upsilon _{\phi _{in}^{-1}}^{(np)}(B\times F_{out}))(\theta )=(\mathrm{M}%
(B))(\theta )\chi _{\phi _{out}^{-1}(F_{out})}(\theta ),  \label{35}
\end{equation}
for all $\theta \in \Theta ,$ $B\in \mathcal{F}_{\Omega },$ $F_{out}\in 
\mathcal{F}_{\Theta _{out}},$\textit{\ where} $\mathrm{M}:\mathcal{F}%
_{\Omega }\rightarrow \frak{B}_{+}\frak{(}\Theta ),$ $\mathrm{M}(\Omega )=%
\mathcal{I}_{_{\frak{B(}\Theta )}},$ \textit{is an outcome generalized
observable on} $(\Theta ,\mathcal{F}_{\Theta })$.
\end{definition}

To see why we call $\Upsilon ^{(np)}$ \textit{non-perturbing}, consider an
experiment $\mathcal{E}$ upon $\mathcal{S},$ represented on $(\Theta _{in},%
\mathcal{F}_{\Theta _{in}})$ by a non-perturbing generalized observable $%
\Upsilon ^{(np)}.$

Let $\mathcal{E}+\mathcal{S}(\pi _{in})$ be an experimental situation where
the initial information state $(\Theta _{in},\mathcal{F}_{\Theta _{in}},[\pi
_{in}])$ is the $\phi _{in}$-image of an information state $(\Theta ,%
\mathcal{F}_{\Theta },[\pi ]),$ that is\footnote{%
See section 3.2.1.}: 
\begin{eqnarray*}
(\Theta _{in},\mathcal{F}_{\Theta _{in}},[\pi _{in}]) &=&\phi _{in}[(\Theta ,%
\mathcal{F}_{\Theta },[\pi ])], \\
\pi _{in} &=&\pi \circ \phi _{in}^{-1}.
\end{eqnarray*}

From (\ref{26}) it follows that, for any $\mathcal{E}+\mathcal{S}(\pi \circ
\phi _{in}^{-1}),$ $\pi \in \mathcal{V}_{(\Theta ,\mathcal{F}_{\Theta })},$
the probability distribution of outcomes is given by 
\begin{eqnarray}
\mu _{\Upsilon ^{(np)}}(B;\pi _{in}) &=&\int_{\Theta _{in}}(\Upsilon
^{(np)}(B\times \Theta _{out}))(\theta _{in})\pi _{in}(d\theta _{in})
\label{36} \\
&=&\int_{\Theta }(\Upsilon _{\phi _{in}^{-1}}^{(np)}(B\times \Theta
_{out}))(\theta )\pi (d\theta )  \notag \\
&=&\int_{\Theta }\mathrm{M}(B)(\theta )\pi (d\theta )=\mu _{\mathrm{M}%
}(B;\pi ),\text{ \ \ \ \ \ \ \ \ \ \ }\forall B\in \mathcal{F}_{\Omega }, 
\notag
\end{eqnarray}
while, for each $B\in \mathcal{F}_{\Omega },$ $\mu _{\Upsilon ^{(np)}}(B;\pi
_{in})\neq 0,$ the conditional posterior information state 
\begin{equation*}
(\Theta _{out},\mathcal{F}_{\Theta _{out}},[\pi _{\Upsilon
^{(np)}}^{out}(\cdot |B;\pi \circ \phi _{in}^{-1})])
\end{equation*}
is represented by 
\begin{eqnarray}
\pi _{\Upsilon ^{(np)}}^{out}(F_{out}|B;\pi \circ \phi _{in}^{-1}) &=&\frac{%
\int_{\Theta _{in}}(\Upsilon ^{(np)}(B\times F_{out}))(\theta _{in})(\pi
\circ \phi _{in}^{-1})(d\theta _{in})}{\mu _{\Upsilon ^{(np)}}(B;\pi _{in})}%
\text{ \ \ \ \ \ \ \ \ }  \label{37} \\
&=&\frac{\int_{\phi _{out}^{-1}(F_{out})}(\mathrm{M}(B))(\theta )\pi
(d\theta )}{\int_{\Theta }(\mathrm{M}(B))(\theta )\pi (d\theta )},\text{ \ \
\ }\forall F_{out}\in \mathcal{F}_{\Theta _{out}}.  \notag
\end{eqnarray}

From (\ref{37}) it follows that, for any $\pi \in \mathcal{V}_{(\Theta ,%
\mathcal{F}_{\Theta })},$ $B\in \mathcal{F}_{\Omega },$ $\mu _{\mathrm{M}%
}(B;\pi )\neq 0,$ we have 
\begin{equation}
\pi _{\Upsilon ^{(np)}}^{out}(F_{out}|B;\pi \circ \phi _{in}^{-1})=\pi
_{\Upsilon _{\phi _{in}^{-1}}^{(np)}}^{out}(\phi _{out}^{-1}(F_{out})|B;\pi
),\text{ \ \ \ \ \ }\forall F_{out}\in \mathcal{F}_{\Theta _{out}},\text{ }
\label{38}
\end{equation}
where 
\begin{equation*}
\pi _{\Upsilon _{\phi _{in}^{-1}}^{(np)}}^{out}(F|B;\pi ):=\frac{\int_{F}(%
\mathrm{M}(B))(\theta )\pi (d\theta )}{\int_{\Theta }(\mathrm{M}(B))(\theta
)\pi (d\theta )},\text{ \ \ \ }\forall F\in \mathcal{F}_{\Theta }.
\end{equation*}

Hence, for any $\pi \in \mathcal{V}_{(\Theta ,\mathcal{F}_{\Theta })},$ $%
B\in \mathcal{F}_{\Omega },$ $\mu _{\mathrm{M}}(B;\pi )\neq 0,$ the
conditional posterior information state is given by 
\begin{equation}
(\Theta _{out},\mathcal{F}_{\Theta _{out}},[\pi _{\Upsilon
^{(np)}}^{out}(\cdot |B;\pi \circ \phi _{in}^{-1})])=\phi _{out}[(\Theta ,%
\mathcal{F}_{\Theta },[\pi _{\Upsilon _{\phi _{in}^{-1}}^{(np)}}^{out}(\cdot
|B;\pi )])].  \label{39}
\end{equation}
In particular, the unconditional posterior information state 
\begin{equation*}
(\Theta _{out},\mathcal{F}_{\Theta _{out}},[\pi _{\Upsilon
^{(np)}}^{out}(\cdot |\Omega ;\pi \circ \phi _{in}^{-1})])=\phi
_{out}[(\Theta ,\mathcal{F}_{\Theta },[\pi ])]
\end{equation*}
is the $\phi _{out}$-image of the initial $\phi _{in}$-preimage state.

Let an initial and an output information state spaces coincide: 
\begin{equation*}
(\Theta _{out},\mathcal{F}_{\Theta _{out}},[\mathcal{V}_{(\Theta _{out},%
\mathcal{F}_{\Theta _{out}})}])=(\Theta _{in},\mathcal{F}_{\Theta _{in}},[%
\mathcal{V}_{(\Theta _{in},\mathcal{F}_{\Theta _{in}})}]),
\end{equation*}
then $\phi _{in}=\phi _{out}=\phi $ and we derive 
\begin{equation}
\pi _{\Upsilon ^{(np)}}^{out}(F_{out}|\Omega ;\pi _{in})])=\pi
_{in}(F_{out}),\text{ \ \ }\forall F_{out}\in \mathcal{F}_{\Theta _{in}}.
\label{40}
\end{equation}
that is, in this case, the unconditional posterior information state
coincides with the initial one.

Suppose that $\mathcal{F}_{\Theta }$ contains all atom subsets of $\Theta $
and let an initial information state of $\mathcal{S}$ in $(\Theta _{in},%
\mathcal{F}_{\Theta _{in}},[\mathcal{V}_{(\Theta _{in},\mathcal{F}_{\Theta
_{in}}}])$ be an $\phi _{in}$-image of the pure information state 
\begin{equation}
(\Theta ,\mathcal{F}_{\Theta },[\delta _{\theta _{0}}]),  \label{41}
\end{equation}
with an arbitrary $\theta _{0}\in \Theta .$ Then from (\ref{37}) it follows
that, for any $B\in \mathcal{F}_{\Omega },$ $\mu _{\mathrm{M}}(B;\pi )\neq
0, $ 
\begin{equation}
\pi _{\Upsilon ^{(np)}}^{out}(F_{out}|B;\delta _{\theta _{0}}\circ \phi
_{in}^{-1})=\chi _{\phi _{out}^{-1}(F_{out})}(\theta _{0}),\text{ \ \ \ \ }%
\forall F_{out}\in \mathcal{F}_{\Theta _{out}}.  \label{42}
\end{equation}
This relation implies that, under any experimental situation $\mathcal{E}%
_{\Upsilon ^{(np)}}+\mathcal{S}(\delta _{\theta _{0}}\circ \phi _{in}^{-1}),$
$\theta _{0}\in \Theta ,$ a conditional posterior information state,
following a single experimental trial, does not depend on an event $B\in 
\mathcal{F}_{\Omega }$, recorded under this trial, and represents the $\phi
_{out}$-image of the initial preimage pure information state (\ref{41}) in $%
(\Theta ,\mathcal{F}_{\Theta },[\mathcal{V}_{(\Theta ,\mathcal{F}_{\Theta
})}])$.

\begin{conclusion}
Under any single experimental trial of an experiment, represented by a
non-perturbing generalized observable, a preimage pure information state on%
\textit{\ }$(\Theta ,\mathcal{F}_{\Theta })$\ is not perturbed. We call this
experiment non-perturbing.
\end{conclusion}

Notice that a\textit{\ non-perturbing experiment may have, in general, a
probabilistic set-up }and represent, for example, a classical measurement
with errors [10].

\subsubsection{Beables}

Analyze now the special case of a non-perturbing generalized observable $%
\widetilde{\Upsilon }^{(np)}$ on $(\Theta _{in},\mathcal{F}_{\Theta _{in}}),$
for which the $\phi _{in}$-preimage (\ref{35}) has the special form: 
\begin{equation}
(\widetilde{\Upsilon }_{\phi _{in}^{-1}}^{(np)}(B\times F_{out}))(\theta
)=\chi _{\varphi ^{-1}(B)\cap \phi _{out}^{-1}(F_{out})}(\theta ),
\label{43}
\end{equation}
for all $\theta \in \Theta ,\ B\in \mathcal{F}_{\Omega },$ $F_{out}\in 
\mathcal{F}_{\Theta _{out}}.$ Here, in addition to the specifications,
introduced in definition 13, a function $\varphi :\Theta \rightarrow \Omega 
\mathrm{\ }$is $\mathcal{F}_{\Theta }/\mathcal{F}_{\Omega }$-measurable.

From definition 7 and proposition 1 it follows that $\widetilde{\Upsilon }%
^{(np)}$is an observable on $(\Theta _{in},\mathcal{F}_{\Theta _{in}})$ and,
hence, describes \textit{an experiment with a} \textit{deterministic set-up. 
}

Due to (\ref{17}), (\ref{29}), for each experimental situation 
\begin{equation*}
\mathcal{E}_{\widetilde{\Upsilon }^{(np)}}+\mathcal{S}(\pi \circ \phi
_{in}^{-1}),\text{ \ \ }\pi \in \mathcal{V}_{(\Theta ,\mathcal{F}_{\Theta
})},
\end{equation*}
the value $(\mathcal{M}_{\widetilde{\Upsilon }^{(np)}}^{pr}(\cdot ))(\pi
\circ \phi _{in}^{-1})$ of the information state instrument is given by 
\begin{equation*}
((\mathcal{M}_{\widetilde{\Upsilon }^{(np)}}^{pr}(B))(\pi \circ \phi
_{in}^{-1}))(F_{out})=\pi (\varphi ^{-1}(B)\cap \phi _{out}^{-1}(F_{out})),
\end{equation*}
for all $B\in \mathcal{F}_{\Omega },\ F_{out}\in \mathcal{F}_{\Theta
_{out}}. $

From (\ref{30}), as well as from (\ref{36}), (\ref{37}), it follows that the
outcome probability law of each experimental situation $\mathcal{E}_{%
\widetilde{\Upsilon }^{(np)}}+\mathcal{S}(\pi \circ \phi _{in}^{-1}),$ $\pi
\in \mathcal{V}_{(\Theta ,\mathcal{F}_{\Theta })},$ is given by the \textit{%
image probability distribution} 
\begin{equation}
\mu _{\widetilde{\Upsilon }^{(np)}}(B;\pi )=\pi (\varphi ^{-1}(B)),\text{ \
\ }\forall B\in \mathcal{F}_{\Omega },  \label{44}
\end{equation}
while, for each $B\in \mathcal{F}_{\Omega },$ $\pi (\varphi ^{-1}(B))\neq 0,$
the conditional posterior state 
\begin{equation*}
(\Theta _{out},\mathcal{F}_{\Theta _{out}},[\pi _{\widetilde{\Upsilon }%
^{(np)}}^{out}(\cdot |B;\pi \circ \phi _{in}^{-1})])
\end{equation*}
is represented by 
\begin{equation}
\pi _{\widetilde{\Upsilon }^{(np)}}^{out}(F_{out}|B;\pi \circ \phi
_{in}^{-1})=\frac{\pi (\varphi ^{-1}(B)\cap \phi _{out}^{-1}(F_{out}))}{\pi
(\varphi ^{-1}(B))},\text{ \ \ \ \ }\forall F_{out}\in \mathcal{F}_{\Theta
_{out}}.  \label{45}
\end{equation}
\smallskip

Suppose that all atom subsets $\{\omega \}$ of $\Omega $ belong to $\mathcal{%
F}_{\Omega }.$

Let an initial information state be an $\phi _{in}$-image of a pure
information state $(\Theta ,\mathcal{F}_{\Theta },[\delta _{\theta _{0}}]).$
In this case, the relations (\ref{44}) and (\ref{45}) imply that, for any
experimental situation 
\begin{equation*}
\mathcal{E}_{\widetilde{\Upsilon }^{(np)}}+\mathcal{S}(\delta _{\theta
_{0}}\circ \phi _{in}^{-1}),\text{ \ \ }\theta _{0}\in \Theta ,
\end{equation*}
the outcome $\omega _{0}=\varphi (\theta _{0})$ is predicted \textit{with
certainty }while the corresponding conditional posterior information state 
\begin{equation*}
(\Theta _{out},\mathcal{F}_{\Theta _{out}},[\pi _{\widetilde{\Upsilon }%
^{(np)}}^{out}(\cdot |B;\delta _{\theta _{0}}\circ \phi _{in}^{-1})]=\phi
_{out}[(\Theta ,\mathcal{F}_{\Theta },[\delta _{\theta _{0}}])],\text{ \ \ \ 
}B\ni \omega _{0},\text{ \ \ }\forall B\in \mathcal{F}_{\Omega },
\end{equation*}
is the $\phi _{out}$-image of an initial $\phi _{in}$-preimage state $%
(\Theta ,\mathcal{F}_{\Theta },[\delta _{\theta _{0}}])$.

Summing up, we introduce the following definition.

\begin{definition}[Beable]
\textit{We call an extended observable\ on} $(\Theta _{in},\mathcal{F}%
_{\Theta _{in}}),$ \textit{with an outcome space} ($\Omega ,\mathcal{F}%
_{\Omega }),$ \textit{a} \textit{\textbf{beable} and, for specificity,
denote it by }$\mathrm{E}^{(be)}$\textit{\ if there exist:\newline
(i)\ a measurable space} $(\Theta ,\mathcal{F}_{\Theta });$\newline
\textit{(ii) an} $\mathcal{F}_{\Theta }/\mathcal{F}_{\Theta _{in}}$-\textit{%
measurable function} $\phi _{in}:\Theta \rightarrow \Theta _{in};$\newline
\textit{(iii)\ an} $\mathcal{F}_{\Theta }/\mathcal{F}_{\Theta _{out}}$-%
\textit{measurable function} $\phi _{out}:\Theta \rightarrow \Theta _{out};$%
\newline
(iiii) an $\mathcal{F}_{\Theta }/\mathcal{F}_{\Omega }$ measurable function $%
\varphi :\Theta \rightarrow \Omega $;\newline
\textit{such that} 
\begin{equation}
(\mathrm{E}_{\phi _{in}^{-1}}^{(be)}(B\times F_{out}))(\theta )=\chi
_{\varphi ^{-1}(B)\cap \phi _{out}^{-1}(F_{out})}(\theta ),  \label{46}
\end{equation}
for all $\theta \in \Theta ,$ $B\in \mathcal{F}_{\Omega },\ F_{out}\in 
\mathcal{F}_{\Theta _{out}}.$
\end{definition}

From this definition it follows that, for the outcome beable $\mathrm{M}%
^{(be)}(B):=\mathrm{E}^{(be)}(B\times \Theta _{out}),$ the $\phi _{in}$%
-preimage on $(\Theta ,\mathcal{F}_{\Theta })$ is given by 
\begin{equation*}
(\mathrm{M}_{\phi _{in}^{-1}}^{(be)}(B))(\theta )=\chi _{\varphi
^{-1}(B)}(\theta ),\text{ \ \ }\forall B\in \mathcal{F}_{\Omega },\text{ \ \ 
}\forall \theta \in \Theta .
\end{equation*}

\textit{A\ beable describes a non-perturbing experiment, with a
deterministic set-up, on an ''errorless'' classical measurement\footnote{%
In the quantum case the term ''measurement'' is understood in a broader
sense and means any experiment upon a quantum system, admitting the
probabilistic description.} of a property }$\varphi $\textit{\ of $\mathcal{S%
}$ on }$(\Theta ,\mathcal{F}_{\Theta })$\textit{, existing objectively
before this measurement. }

\textit{Under an experiment, represented by a beable, the randomness is
caused only by to the uncertainty, encoded in a preimage initial probability
distribution} $\pi \in \mathcal{V}_{(\Theta ,\mathcal{F}_{\Theta })}.$

From the above definition and (\ref{45}) it follows that, under an
experiment, represented by a beable $\mathrm{E}^{(be)}$, a $\phi _{in}$%
-preimage conditional posterior information state in $(\Theta ,\mathcal{F}%
_{\Theta },[\mathcal{V}_{(\Theta ,\mathcal{F}_{\Theta })}])$ is given by 
\begin{equation}
\pi _{\mathrm{E}_{\phi _{out}^{-1}}^{(be)}}^{out}(F_{out}|B;\pi _{in})=\frac{%
\pi (\varphi ^{-1}(B)\cap F_{out})}{\pi (\varphi ^{-1}(B))},  \label{47}
\end{equation}
for any $\pi \in \mathcal{V}_{(\Theta ,\mathcal{F}_{\Theta })},$ and, hence,
is, in general, different from an initial preimage state $(\Theta ,\mathcal{F%
}_{\Theta },[\pi ])$.

\textit{In the frame of Kolmogorov's model, only experiments, represented by
beables, are considered}.

Example 3 and the formulae (\ref{34}), (\ref{45}), (\ref{47}) indicate the
main general reasons of the phenomenon of ''reduction'' of an initial
information state, discussed in 4.2.1.

Specifically, due to (\ref{47}), the reduction of a preimage mixed
information state is inherent even to a classical ''errorless'' measurement
and represents the ''renormalization'' of the initial information on\textit{%
\ }$\mathcal{S},$\textit{\ }conditional on the recorded event under a single
experimental trial.\textit{\ }

The formula (\ref{34})\textit{\ }indicates that, in general, a ''reduction''
of an information\textit{\ }state may be caused not only by the
renormalization of our information on $\mathcal{S},$\ acquired during an
experiment, but also by the ''dynamical'' change of a system state in case
where, with respect to a system, an experiment is perturbing.

\subsection{Equivalence classes of experiments}

We call experiments upon a system $\mathcal{S},$ represented initially by $%
(\Theta _{in},\mathcal{F}_{\Theta _{in}},[\mathcal{V}_{(\Theta _{in},%
\mathcal{F}_{\Theta _{in}})}]),$ \textit{statistically equivalent }and%
\textit{\ }denote by $[\mathrm{M}]$ the corresponding equivalence class, if
these experiments are represented on ($\Theta _{in},\mathcal{F}_{\Theta
_{in}})$ by the same outcome generalized observable $\mathrm{M}$ of $%
\mathcal{S}.$ For all experiments in $[\mathrm{M}]$, the outcome probability
laws on $(\Omega ,\mathcal{F}_{\Omega })$ are identical. In general, an
outcome generalized observable $\mathrm{M}$ may, however, correspond to
different extended generalized observables $\Upsilon $ on $(\Theta _{in},%
\mathcal{F}_{\Theta _{in}}).$

Under all non-destructive experiments upon $\mathcal{S},$ represented by the
same system generalized observable $\mathrm{S}$ on $(\Theta _{in},\mathcal{F}%
_{\Theta _{in}}),$ the unconditional posterior information on $\mathcal{S}$
is represented by the same unconditional posterior information state $%
(\Theta _{out},\mathcal{F}_{\Theta _{out}},[\tau (\cdot ;\pi _{in})]).$ We
denote the corresponding equivalence class of such experiments by $[\mathrm{S%
}].$

We say that non-destructive experiments upon $\mathcal{S}$ are \textit{%
completely information equivalent} if they are represented on $(\Theta _{in},%
\mathcal{F}_{\Theta _{in}})$ by the same extended generalized observable\ $%
\Upsilon $. We denote the corresponding equivalence class by $[\Upsilon ].$
For all experiments in $[\Upsilon ],$ the outcome probability laws on $%
(\Omega ,\mathcal{F}_{\Omega })$ and the families of conditional posterior
information states are identical.

Clearly, $[\Upsilon ]\subseteq \lbrack \mathrm{S}_{\Upsilon }]\cap \lbrack 
\mathrm{M}_{\Upsilon }].$

\subsection{Complete statistical description}

In this section we consider the case where an initial and an output
information state spaces satisfy the condition (\ref{14}). For specificity,
we denote this system by $\mathcal{S}_{\mathrm{V}}.$

Let $\mathcal{E}$ be a non-destructive experiment upon $\mathcal{S}_{\mathrm{%
V}},$ represented on $(\Theta _{in},\mathcal{F}_{\Theta _{in}})$ by a
non-trivial extended generalized $\Upsilon $.

Denote by $\mathrm{V}_{in},$ $\mathrm{V}_{out}$ the corresponding initial
and output Banach spaces in (\ref{14}) and by $\frak{R}_{mean}^{in}$ and $%
\frak{R}_{mean}^{out}$ the corresponding initial and output mean information
state spaces.

Due to definition 4, to an initial information state $(\Theta _{in},\mathcal{%
F}_{\Theta _{in}},[\pi _{in}]),$ the \textit{initial mean information state }%
is defined by 
\begin{equation}
\eta _{in}(\pi _{in}):=\eta _{mean}(\pi _{in})=\int_{\Theta _{in}}\theta
_{in}\pi _{in}(d\theta _{in})\in \frak{R}_{mean}^{in}.  \label{48}
\end{equation}

Immediately after a single trial of an experimental situation $\mathcal{E}+%
\mathcal{S}_{\mathrm{V}}(\pi _{in})$, $\pi _{in}\in \mathcal{V}_{(\Theta
_{in},\mathcal{F}_{\Theta _{in}})}$, where an event $\omega \in B\in 
\mathcal{F}_{\Omega }$ has been recorded, the output mean information state 
\begin{equation}
\eta ^{out}(B;\pi _{in}):=\eta _{mean}(\pi _{\Upsilon }^{out}(\cdot |B;\pi
_{in}))=\int_{\Theta _{out}}\theta _{out}\pi _{\Upsilon }^{out}(d\theta
_{out}|B;\pi _{in})\in \frak{R}_{mean}^{out}  \label{49}
\end{equation}
represents the \textit{conditional statistical average of posterior $%
\mathcal{S}$}-\textit{outcomes}, defined by the conditional probability
distribution (\ref{28}).

\begin{definition}
\textit{We call a conditional statistical average of posterior }$\mathcal{S}$%
-\textit{outcomes }(\ref{49}) \textit{a \textbf{conditional posterior mean
information state} of }$\mathcal{S}_{\mathrm{V}}$ following a single
experimental trial where an event $\omega \in B\in \mathcal{F}_{\Omega }$
has been recorded.
\end{definition}

Due to (\ref{30}) and (\ref{49}), for each $\pi _{in}\in \mathcal{V}%
_{(\Theta _{in},\mathcal{F}_{\Theta _{in}})}$ and each $B\in \mathcal{F}%
_{\Omega },$ $\mu _{\Upsilon }(B,\pi _{in})\neq 0,$ the conditional
posterior mean information state\textit{\ }is defined by 
\begin{equation}
\eta ^{out}(B;\pi _{in})=\frac{\int_{\Theta _{out}}\theta _{out}((\mathcal{M}%
_{\Upsilon }^{pr}(B))(\pi _{in}))(d\theta _{out})}{\mu _{\Upsilon }(B;\pi
_{in})}  \label{50}
\end{equation}
via the notion of the information state instrument $\mathcal{M}_{\Upsilon
}^{pr},$ corresponding to this experiment.

\begin{definition}
\textit{Under the}\textbf{\ complete statistical description}\textit{\ of a
non-destructive experiment }$\mathcal{E}$ \textit{upon a\ system} $\mathcal{S%
}_{\mathrm{V}},$ \textit{we} \textit{mean\ the\ knowledge, for each initial
information state} $(\Theta _{in},\mathcal{F}_{\Theta _{in}},[\pi _{in}]),$%
\textit{\ of the outcome probability law\ }$\mu (\cdot ;\pi _{in})$\textit{\
and the family} $\{\eta ^{out}(B;\pi _{in}):B\in \mathcal{F}_{\Omega }\}$ 
\textit{of all conditional posterior mean information states}.
\end{definition}

To different initial information states, represented by measures $\pi
_{in}\neq \pi _{in}^{\prime },$ but inducing the same initial mean
information states $\eta _{in}(\pi _{in})=\eta _{in}^{\prime }(\pi
_{in}^{\prime }),$ the conditional posterior mean information states,
conditioned by the same recorded event $B\in \mathcal{F}_{\Omega },$ are, in
general, different.

In general, for a non-destructive experiment upon $\mathcal{S}_{\mathrm{V}}$%
, the complete statistical description is less informative than the complete
information description. The latter is due to the fact that the knowledge of
only the initial mean information state $\eta _{in}(\pi _{in})$ does not
allow to make predictions upon all experiments on $\mathcal{S}_{\mathrm{V}},$
described by generalized observables in $\mathcal{G}_{(\Theta _{in},\mathcal{%
F}_{\Theta _{in}})}.$

\subsubsection{Mean information state instrument}

Recall that any complex valued function on a linear space is called a
functional. Denote by $\mathrm{V}_{in}^{\ast }$ the Banach space of all
continuous linear functionals on $\mathrm{V}_{in}.$ Suppose that on $\mathrm{%
V}_{in}$ there exists a\textit{\ continuous linear} functional $\frak{I}%
_{in}\in \mathrm{V}_{in}^{\ast }$ such that 
\begin{equation}
\frak{I}_{in}\{\theta _{in}\}=\mathcal{I}(\theta _{in})=1,\text{ \ \ \ \ \ \
\ }\forall \theta _{in}\in \Theta _{in}\subset \mathrm{V}_{in}.  \label{51}
\end{equation}
In this case, any $\sigma $-additive\footnote{%
Here the convergence in the $\sigma $-additivity condition is in the strong
operator topology in \textrm{V}$_{in}^{\ast }.$} measure 
\begin{equation}
\Pi :\mathcal{F}_{\Lambda }\rightarrow \mathrm{V}_{in}^{\ast }  \label{52}
\end{equation}
on $(\Lambda ,\mathcal{F}_{\Lambda }),$ satisfying the relations 
\begin{equation}
\Pi (\Lambda )=\frak{I}_{in},\text{ \ \ }(\Pi (D))(\eta _{in})\geq 0\text{,
\ \ \ \ \ }\forall \eta _{in}\in \frak{R}_{mean}^{in},\text{ \ }\forall D\in 
\mathcal{F}_{\Lambda },  \label{53}
\end{equation}
defines a generalized observable (extended, or outcome, or system) of a
system $\mathcal{S}_{\mathrm{V}}.$

For short, we refer to this generalized observable as a \textit{linear}
generalized observable and denote it by $\Pi _{lin}.$

The information state instrument, corresponding, due to (\ref{29}), to an
extended \textit{linear} generalized observable $\Upsilon _{lin}$, has the
form 
\begin{equation*}
((\mathcal{M}_{\Upsilon _{lin}}^{pr}(B))(\pi _{in}))(F)=\int_{\Theta
_{in}}(\Upsilon _{lin}(B\times F))(\theta _{in})\pi _{in}(d\theta
_{in})=(\Upsilon _{lin}(B\times F))(\eta _{in}),
\end{equation*}
for all $\pi _{in}\in \mathcal{V}_{(\Theta _{in},\mathcal{F}_{\Theta
_{in}})},$ $B\in \mathcal{F}_{\Omega },$ $F\in \mathcal{F}_{\Theta _{out}},$
and, hence, \textit{depends only on the initial mean information state} $%
\eta _{in}=\eta _{mean}(\pi _{in})$ but not on an initial information state,
represented by $\pi _{in}.$

From (\ref{30}), (\ref{50}) it follows that, to each experimental situation $%
\mathcal{E}_{\Upsilon _{lin}}+\mathcal{S}(\pi _{in})$, $\pi _{in}\in 
\mathcal{V}_{(\Theta _{in},\mathcal{F}_{\Theta _{in}})},$ the probability
distribution of outcomes in $(\Omega ,\mathcal{F}_{\Omega })$ 
\begin{equation*}
\mu _{\Upsilon _{lin}}(B;\pi _{in})=(\Upsilon _{lin}(B\times \Theta
_{out}))(\eta _{in})=(\mathrm{M}_{\Upsilon _{lin}}(B))(\eta _{in}):=%
\widetilde{\mu }_{\Upsilon _{lin}}(B,\eta _{in}),\text{ \ \ }\forall B\in 
\mathcal{F}_{\Omega }\text{\ }
\end{equation*}
and the conditional posterior mean information states 
\begin{equation}
\eta ^{out}(B;\pi _{in})=\frac{\int_{\Theta _{out}}\theta _{out}(\Upsilon
_{lin}(B\times d\theta _{out}))(\eta _{in})}{\mu _{\Upsilon _{lin}}(B,\eta
_{in})}:=\widetilde{\eta }^{out}(B;\eta _{in}),\text{ \ \ \ \ }\forall B\in 
\mathcal{F}_{\Omega }  \label{54}
\end{equation}
\textit{depend} \textit{only} \textit{on the initial mean information state} 
$\eta _{in}.$

Denote in (\ref{54}) 
\begin{equation}
(\mathcal{M}_{\Upsilon _{lin}}^{st}(B))(\eta _{in}):=\int_{\Theta
_{out}}\theta _{out}(\Upsilon _{lin}(B\times d\theta _{out}))(\eta _{in})\in 
\mathrm{V}_{out},  \label{55}
\end{equation}
$\forall \eta _{in}\in \frak{R}_{mean}^{in},$ $\forall B\in \mathcal{F}%
_{\Omega }.$

\begin{definition}
\textit{We call the mapping }$\mathit{(}\mathcal{M}_{\Upsilon
_{lin}}^{st}(\cdot ))(\cdot ):\mathcal{F}_{\Omega }\times \frak{R}%
_{mean}^{in}\rightarrow \mathrm{V}_{out},$ \textit{defined by an extended
linear generalized observable }$\Upsilon _{lin}$ \textit{through the relation%
} 
\begin{equation}
(\mathcal{M}_{\Upsilon _{lin}}^{st}(B))(\eta _{in}):=\int_{\Theta
_{out}}\theta _{out}(\Upsilon _{lin}(B\times d\theta _{out}))(\eta _{in}),
\label{56}
\end{equation}
for all $B\in \mathcal{F}_{\Omega },$ $\eta _{in}\in \frak{R}_{mean}^{in},$ 
\textit{a }\textbf{mean information state instrument.}
\end{definition}

For each $\eta _{in}\in \frak{R}_{mean}^{in}$,\ the mapping\textit{\ }$(%
\mathcal{M}_{_{\Upsilon _{lin}}}^{(st)}(\cdot ))(\eta _{in})$ is a\textit{\ }%
$\sigma $-additive measure on $(\Omega ,\mathcal{F}_{\Omega })$ with values
in\textit{\ }$\mathrm{V}_{out}$.

\begin{remark}
To a linear extended generalized observable $\Upsilon _{lin}$ of $\mathcal{S}%
_{\mathrm{V}},$ there corresponds a unique mean information state instrument 
\textit{though not vice versa.}
\end{remark}

\begin{remark}
Notice that the notion of a mean information state instrument appears:%
\textit{\ }\newline
(i) \textit{only\ }for a system\textit{\ }$\mathcal{S}_{\mathrm{V}},$\textit{%
\ }with the definite type\textit{\ }(\ref{14})\ of measurable spaces\textit{%
\ }$(\Theta _{in},$\textit{$\mathcal{F}$}$_{in})$\textit{\ }and\textit{\ }$%
(\Theta _{out},$\textit{$\mathcal{F}$}$_{out});$\newline
(ii)\textit{\ only\ }in case where on a Banach space $\mathrm{V}_{in}$ there
exists a continuous linear functional, satisfying (\ref{51});\newline
(iii)\textit{\ only\ }to a non-destructive experiment\textit{\ $\mathcal{E}$ 
}upon\textit{\ }$\mathcal{S}_{\mathrm{V}}$\textit{, }represented by a linear
generalized observable\emph{\ }$\Upsilon _{lin}.$
\end{remark}

From (\ref{54}) and (\ref{56}) it follows that, under an non-destructive
experiment $\mathcal{E}\in \lbrack \Upsilon _{lin}],$ to any initial mean
information state $\eta _{in}$ of $\mathcal{S}$ and any $B\in \mathcal{F}%
_{\Omega },\widetilde{\mu }_{\mathrm{\Upsilon }_{lin}}(B;\eta _{in})\neq 0$,
the conditional posterior mean information state depends only on $\eta _{in}$
and is given by 
\begin{equation}
\widetilde{\eta }^{out}(B;\eta _{in})=\frac{(\mathcal{M}_{\Upsilon
_{lin}}^{st}(B))(\eta _{in})}{\widetilde{\mu }_{\Upsilon _{lin}}(B;\eta
_{in})}.  \label{57}
\end{equation}

In the most general settings, the conditional change 
\begin{equation*}
\eta _{in}\mapsto \widetilde{\eta }^{out}(B;\eta _{in})
\end{equation*}
represents the\textit{\ phenomenon of ''reduction'' of a mean information
state, induced by the reduction of an information state of}\emph{\ }$%
\mathcal{S}_{\mathrm{V}}$ (see section 4.2.1)\textit{.}

If, further, on the Banach space $\mathrm{V}_{out}$ there also exists a 
\textit{continuous linear} functional $\frak{J}_{out}:\mathrm{V}%
_{out}\rightarrow \mathbb{C}$ such that 
\begin{equation*}
\frak{I}_{out}\{\eta _{out}\}=1,\text{ \ \ }\forall \eta _{out}\in \frak{R}%
_{mean}^{out},
\end{equation*}
then 
\begin{equation}
\widetilde{\mu }_{\Upsilon _{lin}}(B;\eta _{in})=\frak{J}_{out}\{(\mathcal{M}%
_{\Upsilon _{lin}}^{st}(B))(\eta _{in})\},\text{ \ \ \ }\forall \eta
_{in}\in \frak{R}_{mean}^{out},\text{ \ \ \ }\forall B\in \mathcal{F}%
_{\Omega },  \label{58}
\end{equation}
and \textit{in this case the mean information state instrument\ }$\mathcal{M}%
_{\Upsilon _{lin}}^{st}$ \textit{gives the complete statistical description }%
of the corresponding equivalence class $[\Upsilon _{lin}]$ of experiments
upon $\mathcal{S}_{\mathrm{V}}$ \textit{although not the complete
information description}\footnote{%
Under the description of quantum measurements, this point was first
indicated in [17], where we, in particular, introduced the concept of
stochastic realizations of a quantum instrument. Different equivalence
classes of stochastic realizations of the same quantum instrument represent
different experiments.}.

\section{Probabilistic and statistical models}

As we discussed in section 4.3.1, in Kolmogorov's model only the description
of non-perturbing experiments is considered and the random character of
predictions is caused only by the uncertainty, encoded in a probability
distribution $\pi _{in}.$ The uncertainty, encoded in elements of a set $%
\Theta _{in},$ as well as the description of perturbing experiments, are not
analyzed.

\begin{definition}
\textit{We\ call a pair }$(\mathcal{U}_{\mathcal{S}},$ $\mathcal{G}_{_{%
\mathcal{S}}}^{ext})$\textit{\ where:\smallskip }\newline
(i) $\mathcal{U}_{S}\frak{\ }$\textit{is a specified family of initial and
output information state spaces of} $\mathcal{S};$\newline
(ii) $\mathcal{G}_{_{\mathcal{S}}}^{ext}$ \textit{is a specified family of
non-trivial extended generalized observables\ on initial information state
spaces in }$\mathcal{U}_{S};\smallskip $\newline
\textit{a\textbf{\ probabilistic model} for the description of
non-destructive experiments upon }$\mathcal{S}$\textit{.}
\end{definition}

A probabilistic model gives the complete information description of all
non-destructive experiments upon $\mathcal{S},$ represented by extended
generalized observable in $\mathcal{G}_{_{\mathcal{S}}}^{ext}.$

\begin{definition}
\textit{We call a pair }$(\mathcal{U}_{S},$ $\mathcal{G}_{_{\mathcal{S}%
}}^{outcome})$ \textit{where}:\smallskip \newline
(i) $\mathcal{U}_{S}$ \textit{is a specified family of initial information
state spaces of} $\mathcal{S}$;\newline
(ii) $\mathcal{G}_{_{\mathcal{S}}}^{outcome}$ \textit{is a specified family
of non-trivial outcome generalized observables on initial information state
spaces in }$\mathcal{U}_{S};$\smallskip \newline
\textit{a \textbf{statistical model} for the description of experiments upon 
}$\mathcal{S}$\textit{.}
\end{definition}

In case of a non-destructive experiment, the concept of a statistical model
is less informative and gives predictions only\ on outcome probability laws.
We call this type of description of an experiment \textit{statistical}.

In natural sciences, for a system of a concrete type, the specification of a
family $\mathcal{U}_{\mathcal{S}}$ of information state spaces and a family $%
\mathcal{G}_{\mathcal{S}}$ of allowed non-trivial generalized observables on
these information state spaces must be based on the fundamental laws,
describing this concrete type of a system.

As the following proposition shows, in general, there is a correspondence
between families $\mathcal{U}_{\mathcal{S}}$ and $\mathcal{G}_{\mathcal{S}}.$

Consider a system $\widetilde{\mathcal{S}}_{\mathrm{V}},$ specified by the
following conditions:\newline
(i) all possible information state spaces in a family $\mathcal{U}_{%
\widetilde{\mathcal{S}}_{\mathrm{V}}}$ satisfy the condition (\ref{14});%
\newline
(ii) on each Banach space \textrm{V }in (\ref{14}), corresponding to an
information state space in $\mathcal{U}_{\widetilde{\mathcal{S}}_{\mathrm{V}%
}},$ there exists a continuous linear functional $\frak{J}_{\mathrm{V}%
}\{\cdot \}$;\newline
(iii) for each Banach space \textrm{V} in (\ref{14}),\textrm{\ }a mean
information state space $\frak{R}_{mean}$ is such that any element $u\in $%
\textrm{V} admits a representation $u=\beta _{1}\eta _{1}-\beta _{2}\eta
_{2},$ with $||u||_{\mathrm{V}}=\inf (\beta _{1}+\beta _{2}),$ where $\eta
_{1},\eta _{2}\in \frak{R}_{mean},$ $\beta _{1},\beta _{2}\geq 0$. The
latter relation implies, in particular, that, for this system, in (\ref{14}%
), $C\geq 1.$

\begin{proposition}
\textit{If, under a non-destructive experiment }$\mathcal{E}$ \textit{upon} $%
\widetilde{\mathcal{S}}_{\mathrm{V}}$, \textit{for each} $\pi _{in}\in 
\mathcal{V}_{(\Theta _{in},\mathcal{F}_{\Theta _{in}})},$ \textit{a} \textit{%
probability distribution} $\gamma (\cdot ;\pi _{in})$ \textit{(extended, or
outcome, or system)} \textit{depends only on the corresponding initial mean
state }$\eta _{in}(\pi _{in}),$ \textit{but not on} $\pi _{in},$ \textit{%
that is}, $\gamma (\cdot ;\pi _{in})=\widetilde{\gamma }(\cdot ;\eta _{in}),$%
\textit{\ for each} $\pi _{in}\in \mathcal{V}_{(\Theta _{in},\mathcal{F}%
_{\Theta _{in}})},$ $\eta _{in}(\pi _{in})=\eta _{in},$ \textit{then } 
\begin{equation*}
\widetilde{\gamma }(\cdot ;\alpha _{1}\eta _{in}^{(1)}+\alpha _{2}\eta
_{in}^{(2)})=\alpha _{1}\widetilde{\gamma }(\cdot ;\eta _{in}^{(1)})+\alpha
_{2}\widetilde{\gamma }(\cdot ;\eta _{in}^{(2)}),
\end{equation*}
\textit{for any }$\eta _{in}^{(1)},\eta _{in}^{(2)}\in \frak{R}_{mean}^{in},$
$\alpha _{1},\alpha _{2}\geq 0,$ $\alpha _{1}+\alpha _{2}=1,$ \textit{and
this situation is possible iff }$\mathcal{E}$\textit{\ is represented on }$%
(\Theta _{in},\mathcal{F}_{\Theta _{in}})$ \textit{by a linear generalized
observable} $\mathrm{\Pi }_{lin}$ \textit{and} 
\begin{equation*}
\widetilde{\gamma }(\cdot ;\eta _{in})=(\mathrm{\Pi }_{lin}(\cdot ))(\eta
_{in}),\text{ \ \ \ }\forall \eta _{in}\in \frak{R}_{mean}^{in}.
\end{equation*}
\end{proposition}

\begin{remark}
If, under all really performed experiments upon a concrete system $%
\widetilde{\mathcal{S}}_{\mathrm{V}}$, probability distributions of outcomes
depend only on initial mean information states, then, due to this
proposition, for the description of experiments upon $\widetilde{\mathcal{S}}%
_{\mathrm{V}}$, only linear generalized observables $\Pi _{lin}$ (extended,
or outcome, or system) of $\widetilde{\mathcal{S}}_{\mathrm{V}}$ are
allowed.\smallskip
\end{remark}

Let, for a system $\widetilde{\mathcal{S}}_{\mathrm{V}},$ only linear
generalized observables be allowed and 
\begin{equation*}
(\mathcal{U}_{\widetilde{\mathcal{S}}_{\mathrm{V}}},\mathcal{G}_{\widetilde{%
\mathcal{S}}_{\mathrm{V}}}^{ext\text{\&}lin})
\end{equation*}
be a probabilistic model for the description of non-destructive experiments
upon a system $\widetilde{\mathcal{S}}_{\mathrm{V}}$.

If we are interested only in the \textit{complete statistical description}
(section 4.5), we can equivalently replace $\mathcal{U}_{\widetilde{\mathcal{%
S}}_{\mathrm{V}}}$ by the family\textit{\ }$\frak{R}_{mean}(\mathcal{U}_{%
\widetilde{\mathcal{S}}_{\mathrm{V}}})$\textit{\ }of all mean information
state spaces, corresponding to the information state spaces in\ $\mathcal{U}%
_{\widetilde{\mathcal{S}}_{\mathrm{V}}}.$ We call the pair 
\begin{equation*}
(\frak{R}_{mean}(\mathcal{U}_{\widetilde{\mathcal{S}}_{\mathrm{V}}}),%
\mathcal{G}_{\widetilde{\mathcal{S}}_{\mathrm{V}}}^{ext\text{\&}lin}),
\end{equation*}
the \textit{reduced probabilistic model}.

For the reduced probabilistic model, the complete statistical description is
represented by the notion of a mean information state instrument,
introduced, in the most general settings, in section 4.5.1.

Further, the statistical description of experiments upon a system $%
\widetilde{\mathcal{S}}_{\mathrm{V}}$, in the frame of the statistical model 
$(\mathcal{U}_{\widetilde{\mathcal{S}}_{\mathrm{V}}},\mathcal{G}_{\widetilde{%
\mathcal{S}}_{\mathrm{V}}}^{outcome\text{\&}lin})$ coincides with the
statistical description in the frame of the corresponding reduced
statistical model.

\subsection{Reducible models}

Let $(\mathcal{U}_{\mathcal{S}},\mathcal{G}_{\mathcal{S}})$ and $(\widetilde{%
\mathcal{U}}_{\mathcal{S}},\widetilde{\mathcal{G}}_{\mathcal{S}})$ be two
different (probabilistic, or statistical) models for the description of
experiments upon a system $\mathcal{S}.$

We say that a model $(\widetilde{\mathcal{U}}_{\mathcal{S}},\widetilde{%
\mathcal{G}}_{\mathcal{S}})$ is reducible to a model $(\mathcal{U}_{\mathcal{%
S}},\mathcal{G}_{\mathcal{S}})$ if:\medskip

(i) any information state space in $\widetilde{\mathcal{U}}_{\mathcal{S}}$
is induced by an information state space in $\mathcal{U}_{\mathcal{S}%
};\smallskip $

(ii) any generalized observable in $\widetilde{\mathcal{G}}_{\mathcal{S}}$
and any system generalized observable in $\mathcal{G}_{\mathcal{S}},$
generate by (\ref{22}) a generalized observable in $\mathcal{G}_{\mathcal{S}%
}.$

\subsection{Kolmogorov's model}

From our presentation\ in section 4.3 it follows that, in our framework,
Kolmogorov's probabilistic model for the description of experiments upon $%
\mathcal{S}$ can be specified as the pair 
\begin{equation*}
(\mathcal{U}_{\mathcal{S}},\mathcal{G}_{_{\mathcal{S}}}^{be}),
\end{equation*}
for which there exists an information state space ($\Theta ,\mathcal{F}%
_{\Theta },[\mathcal{V}_{(\Theta ,\mathcal{F}_{\Theta })}])$ such that all
observables in $\mathcal{G}_{_{\mathcal{S}}}^{be}$ are beables with the
corresponding preimage observables (\ref{46}), defined on $(\Theta ,\mathcal{%
F}_{\Theta }).$

\textit{In conventional probability theory, the existence of an
''underlying'' information state space }$(\Theta ,\mathcal{F}_{\Theta },[%
\mathcal{V}_{(\Theta ,\mathcal{F}_{\Theta })}])$\textit{\ is postulated}.

In case of Kolmogorov's model, we call $(\Theta ,\mathcal{F}_{\Theta },[%
\mathcal{V}_{(\Theta ,\mathcal{F}_{\Theta })}])$ as \textit{a probability
state space} and any information state in this space as a \textit{%
probability state}.

Kolmogorov's statistical model can be specified as the pair $(\mathcal{U}_{%
\mathcal{S}},\mathcal{G}_{_{\mathcal{S}}}^{outcome\text{\&}be}),$ with $%
\mathcal{G}_{_{\mathcal{S}}}^{outcome\text{\&}be}$ representing the family
of outcome beables with preimages on $(\Theta ,\mathcal{F}_{\Theta }).$

\section{Description of quantum measurements}

In quantum theory, a system is described in terms of a separable complex
Hilbert space $\mathcal{H},$ in general, infinite dimensional.

Denote by $\mathcal{L(H)}$ and $\mathcal{T(H)},$ the Banach spaces of
bounded linear operators and trace class operators on $\mathcal{H},$
respectively. Let $\mathcal{L}^{(+)}\mathcal{(H)\subset L(H)}$ and $\mathcal{%
T}^{(+)}\mathcal{(H)\subset T(H)}$ be the sets of all non-negative operators
in the corresponding Banach spaces.

As we mentioned, in the most general settings in section 5,\ for the
description of non-destructive experiments upon a system, \textit{we
distinguish between a statistical model and a probabilistic model}. With
respect to a system, a probabilistic model is more detailed and includes the
specification of a system conditional posterior probability state following
each trial of this experiment.

In this section we introduce the probabilistic model, the reduced
probabilistic model and the statistical model for the description of
experiments upon a quantum system and prove the corresponding statements.

Under \textit{a generalized quantum measurement,} \textit{we further mean any%
} \textit{experiment upon a quantum system which admits the probabilistic
description and results in imprints in the classical world of any most
general possible nature.}

\subsection{Quantum probabilistic model}

\subsubsection{Information states, mean information states}

Consider a quantum system $\mathcal{S}_{q}$, described in terms of $\mathcal{%
H}$. Denote by $\mathcal{O}_{\mathcal{H}}=\{\psi \in \mathcal{H}:||\psi
||_{_{\mathcal{H}}}=1\}$ the unit sphere in $\mathcal{H}$. Let 
\begin{equation*}
\mathcal{P}_{1}(\mathcal{H)}=\{p\in \mathcal{T}^{(+)}\mathcal{(H)}:p=|\psi
\rangle \langle \psi |,\text{ \ }\forall \psi \in \mathcal{O}_{\mathcal{H}}\}
\end{equation*}
be the set of all one-dimensional projections\footnote{%
Called also as pure density operators.} on $\mathcal{H}$ and $\mathcal{B}_{%
\mathcal{P}_{1}\mathcal{(H)}}$ be the trace on $\mathcal{P}_{1}(\mathcal{H)}$
of the Borel $\sigma $-algebra on $\mathcal{T(H)}.$\ For our further
consideration, we also denote by $R:\mathcal{O}_{\mathcal{H}}\rightarrow 
\mathcal{P}_{1}(\mathcal{H)}$ the surjective mapping $\psi \mapsto p=R(\psi
):=|\psi \rangle \langle \psi |.$

\begin{definition}
\textit{For a quantum system, described in terms\ of\ }$\mathcal{H}$, 
\textit{introduce an} \textbf{information state} \textit{by} 
\begin{equation*}
(\mathcal{P}_{1}(\mathcal{H}),\mathcal{B}_{\mathcal{P}_{1}\mathcal{(H)}%
},[\pi ])
\end{equation*}
for any $\pi \in \mathcal{V}_{(\mathcal{P}_{1}(\mathcal{H)},\mathcal{B}_{%
\mathcal{P}_{1}\mathcal{(H)}})},$ \textit{and an}\textbf{\ information state
space} \textit{by} 
\begin{equation*}
(\mathcal{P}_{1}(\mathcal{H}),\mathcal{B}_{\mathcal{P}_{1}\mathcal{(H)}},[%
\mathcal{V}_{(\mathcal{P}_{1}(\mathcal{H)},\mathcal{B}_{\mathcal{P}_{1}%
\mathcal{(H)}})}]).
\end{equation*}
\end{definition}

In the quantum case, the measurable space $(\mathcal{P}_{1}(\mathcal{H)},%
\mathcal{B}_{\mathcal{P}_{1}\mathcal{(H)}})$ satisfies the condition (\ref
{14}), with \textrm{V} being the Banach space $\mathcal{T}_{s}\mathcal{(H)}$
of self-adjoint trace class operators on $\mathcal{H}.$

For any $\pi \in \mathcal{V}_{(\mathcal{P}_{1}(\mathcal{H)},\mathcal{B}_{%
\mathcal{P}_{1}\mathcal{(H)}})},$ a \textit{quantum} \textit{mean
information state }is given by 
\begin{equation*}
\eta _{mean}(\pi )=\int_{\mathcal{P}_{1}(\mathcal{H})}p\pi (dp)=\int_{%
\mathcal{O}_{\mathcal{H}}}|\psi \rangle \langle \psi |\text{ }\pi ^{\prime
}(d\psi )=\rho ,
\end{equation*}
and represents \textit{a density operator }$\rho $ \textit{on }$\mathcal{H}.$
Here, to each $\pi \in \mathcal{V}_{(\mathcal{P}_{1}(\mathcal{H)},\mathcal{B}%
_{\mathcal{P}_{1}\mathcal{(H)}})},$ the measure $\pi ^{\prime }$ is defined
uniquely by the relation 
\begin{equation*}
\pi ^{\prime }(R^{-1}(F))=\pi (F),\text{ \ \ \ \ }\forall F\in \mathcal{B}_{%
\mathcal{P}_{1}\mathcal{(H)}},
\end{equation*}
and is a normalized $\sigma $-additive positive real valued measure on the $%
\sigma $-algebra $\mathcal{F}_{\mathcal{O}_{\mathcal{H}}}=\{R^{-1}(F):F\in 
\mathcal{B}_{\mathcal{P}_{1}\mathcal{(H)}}\}$ on $\mathcal{O}_{\mathcal{H}}.$

The convex linear set $\frak{R}_{mean}$ of all quantum mean information
states coincides with the set 
\begin{equation*}
\mathcal{R}_{\mathcal{H}}=\{\rho \in \mathcal{T}^{(+)}\mathcal{(H)}:||\rho
||_{_{\mathcal{T(H)}}}=\mathrm{tr}\{\rho \}=1\}
\end{equation*}
of all density operators on $\mathcal{H}$.

Since the $\sigma $-algebra $\mathcal{B}_{\mathcal{P}_{1}\mathcal{(H)}}$
contains all atom subsets $\{p\}$ of $\mathcal{P}_{1}(\mathcal{H)}$ and $%
\mathcal{P}_{1}(\mathcal{H)}$ is the set of all extreme elements in $%
\mathcal{R}_{\mathcal{H}}$, to each pure information state of a quantum
system there corresponds the unique pure mean information state and vice
versa.

According to the terminology, used in quantum theory, \textit{we further
refer to a quantum mean information state, represented by a density operator}
$\rho $ \textit{on} $\mathcal{H},$ \textit{as a quantum state, pure or mixed.%
}

We further denote by 
\begin{equation}
\mathcal{U}_{\mathcal{K}}=(\mathcal{P}_{1}(\mathcal{K}),\mathcal{B}_{%
\mathcal{P}_{1}\mathcal{(K)}},[\mathcal{V}_{(\mathcal{P}_{1}(\mathcal{K)},%
\mathcal{B}_{\mathcal{P}_{1}\mathcal{(K)}})}])  \label{59}
\end{equation}
a possible information state space of a quantum system $\mathcal{S}_{q}$.
Here $\mathcal{K}$ is a separable complex Hilbert space.

\subsubsection{Generalized observables}

Recall that a \textit{linear} functional $\Phi (\cdot ):\mathcal{T(H)}%
\rightarrow \mathbb{C}$ is called non-negative if $\Phi (T)\geq 0$ whenever $%
T\in \mathcal{T}^{(+)}\mathcal{(H)}$. Any non-negative linear functional on $%
\mathcal{T(H)}$ is bounded (equivalently, continuous), that is, there exists
some $C\mathbb{>}0$ such that $|\Phi (T)|\leq C||T||_{\mathcal{T(H)}}.$

Denote by $\mathcal{T}^{\ast }\mathcal{(H)}$ the Banach space of bounded
linear functionals on $\mathcal{T(H)}$ and by $\mathcal{T}_{+}^{\ast }%
\mathcal{(H)\subset T}^{\ast }\mathcal{(H)}$ the set of all non-negative
bounded linear functionals on $\mathcal{T(H)}.$

In the quantum case, the conditions (i)-(iii), specified in section 4.5.1,
are valid. In particular, the mapping $\mathrm{tr}\{\cdot \}:\mathcal{T(H)}%
\rightarrow \mathbb{C}$ represents a unique positive continuous linear
functional $\frak{J}\{\cdot \}$ on $\mathcal{T(H)},$ satisfying (\ref{51}).
Hence, all items, introduced in section 4.5.1 in the most general settings,
are applicable to the quantum case.

Any $\sigma $-additive measure 
\begin{equation*}
\Pi _{lin}^{(q)}:\mathcal{F}_{\Lambda }\rightarrow \mathcal{T}_{+}^{\ast }%
\mathcal{(H)},\ \ \ (\Pi _{lin}^{(q)}(\Lambda ))(T)=\mathrm{tr}\{T\},\text{
\ \ }\forall T\in \mathcal{T(H)},
\end{equation*}
on a measurable space $(\Lambda ,\mathcal{F}_{\Lambda }),$ with values $\Pi
_{lin}^{(q)}(B),$ $B\in \mathcal{F}_{\Lambda },$ that are non-negative
continuous linear functionals on $\mathcal{T(H)}$, satisfies conditions (\ref
{53}), and, hence, represents a linear generalized observable of a quantum
system.

Due to the linear isometric isomorphism between $\mathcal{L(H)}$ and $%
\mathcal{T}^{\ast }\mathcal{(H}),$ to each non-negative continuous linear
functional $\Pi _{lin}^{(q)}(B),$ $B\in \mathcal{F}_{\Lambda },$ there
corresponds the uniquely defined non-negative bounded linear operator $%
A_{\Pi _{lin}}(B),$ $B\in \mathcal{F}_{\Lambda },$ and vice versa, such that 
\begin{equation*}
(\Pi _{lin}^{(q)}((B))(T)=\mathrm{tr}\{TA_{\Pi _{lin}^{(q)}}(B)\},\text{ \ \
\ }\forall T\in \mathcal{T(H)}.
\end{equation*}
Since $\Pi _{lin}^{(q)}:\mathcal{F}_{\Lambda }\rightarrow \mathcal{T}%
_{+}^{\ast }\mathcal{(H)}$ is a $\sigma $-additive measure and 
\begin{equation*}
\mathrm{tr}\{\rho A_{\Pi _{lin}^{(q)}}(\Lambda )\}=1,\text{ \ \ \ \ }\forall
\rho \in \mathcal{R}_{\mathcal{H}},
\end{equation*}
each $A_{\Pi _{lin}^{(q)}}(B)\in \mathcal{L}^{(+)}\mathcal{(H)},$ $B\in 
\mathcal{F}_{\Lambda },$ represents a value of the normalized positive
operator valued measure 
\begin{equation*}
A_{\Pi _{lin}^{(q)}}:\mathcal{F}_{\Lambda }\rightarrow \mathcal{L}^{(+)}%
\mathcal{(H)},\text{ \ \ \ }A_{\Pi _{lin}^{(q)}}(\Lambda )=I_{\mathcal{H}},
\end{equation*}
on $(\Lambda ,\mathcal{F}_{\Lambda })$\footnote{%
Measures $\Pi _{lin}^{q}$ and $A_{\Pi _{lin}^{q}}$ are bounded in the sense
that, $||\Pi ^{q}((B)||_{\mathcal{T}^{\ast }\mathcal{(H)}}\leq 1$ and $%
||A_{\Pi _{lin}^{{}}}(B)||_{\mathcal{L(H})}\leq 1,$ for all $B\in \mathcal{F}%
_{\Lambda }.$}, which is $\sigma $-additive in the strong operator topology
in $\mathcal{L(H)}.\smallskip $

\textit{This implies that, to each linear generalized observable }$\Pi
_{lin}^{(q)}$\textit{\ (extended, or outcome, or system) of a quantum
system, with an outcome space }\emph{(}$\Lambda ,\mathcal{F}_{\Lambda }),$ 
\textit{there corresponds the unique (extended, or outcome, or system)\
normalized }$\sigma $\textit{-additive positive operator valued measure }$%
A_{\Pi _{lin}^{q}}$\textit{\ on }$\mathit{(}\Lambda ,\mathcal{F}_{\Lambda })$
\textit{and vice versa. }

\textit{A} \textit{projection valued measure}\emph{\ }$P$\emph{\ }\textit{on}%
\emph{\ (}$\Lambda ,\mathcal{F}_{\Lambda })$\emph{\ }\textit{represents a
quantum measurement with a deterministic set-up.\smallskip }

Furthermore, real experimental situations show that, under any quantum
measurement, the outcome probability law $\mu (B;\rho _{in})$ on ($\Omega ,%
\mathcal{F}_{\Omega })$ depends only on an initial quantum state $\rho _{in}$
of $\mathcal{S}_{q}$ but not on an initial probability distribution $\pi
_{in}$ on $(\mathcal{P}_{1}(\mathcal{H)},\mathcal{B}_{\mathcal{P}_{1}%
\mathcal{(H)}}).$

Due to proposition 4, this implies that,\textit{\ in the quantum case}, 
\textit{only linear outcome generalized observables }$\mathrm{M}_{lin}^{(q)}$
\textit{are allowed and} 
\begin{equation}
\mu (B;\rho _{in})=(\mathrm{M}_{lin}^{(q)}(B))(\rho _{in})=\mathrm{tr}\{\rho
_{in}M(B)\},  \label{60}
\end{equation}
for all $\rho _{in}\in \mathcal{R}_{\mathcal{H}}$, $B\in \mathcal{F}_{\Omega
}.$ A normalized positive operator valued measure 
\begin{equation*}
M:\mathcal{F}_{\Omega }\rightarrow \mathcal{L}^{(+)}\mathcal{(H)},\text{ \ \
\ \ \ \ \ }M(\Omega )=I_{\mathcal{H}},
\end{equation*}
is usually called in quantum measurement theory as a \textit{probability
operator valued measure }or a \textit{POV measure}, for short.

Furthermore, in quantum measurement theory\footnote{%
See, for example, in [16] and [17-19].} \textit{it is postulated }that,
under a non-destructive quantum measurement, with outcomes in $(\Omega ,%
\mathcal{F}_{\Omega })$, for any $B\in \mathcal{F}_{\Omega },$ the
conditional change of an initial quantum state 
\begin{equation*}
\rho _{in}\mapsto \rho ^{out}(B;\rho _{in})=\frac{(\mathcal{M}(B))(\rho
_{in})}{\mu (B;\rho _{in}\}},
\end{equation*}
following a single experimental trial, is described by a positive\textit{\ }%
bounded\textit{\ linear }mapping $\mathcal{M}$ on $\mathcal{F}_{\Omega
}\times \mathcal{T}_{\mathcal{H}}$. This assumption is usually justified by
a unitary linear ''dynamics'' of a quantum state of the extended quantum
system, which includes the quantum environment of a measuring device.

In our terminology, the latter assumption implies:

\textit{For the description of experiments upon a quantum system}, \textit{%
only linear (extended, system and outcome) generalized observables are
allowed. Hence, with respect to a system, any quantum measurement is
perturbing.}

Summing up, the probabilistic model for the description of non-destructive
quantum measurements upon a quantum system $\mathcal{S}_{q}$ is given by the
pair 
\begin{equation*}
(\mathcal{U}_{\mathcal{S}q},\mathcal{A}_{\mathcal{S}_{q}}^{ext})
\end{equation*}
where:\medskip \newline
(a) $\mathcal{U}_{\mathcal{S}q}=\{\mathcal{U}_{\mathcal{K}_{\gamma
}}:\forall \gamma \in \Gamma \},$ with each $\mathcal{U}_{\mathcal{K}%
_{\gamma }}$ defined by (\ref{41}), being the family of an input and output
information state spaces of $\mathcal{S}_{q};\smallskip $\newline
(b) $\mathcal{A}_{\mathcal{S}q}^{ext}$ is the family of all non-trivial
extended normalized $\sigma $-additive positive operator valued measures 
\begin{equation*}
A:\mathcal{F}_{\Omega }\otimes \mathcal{B}_{\mathcal{P}_{1}(\mathcal{K}%
_{\gamma _{2}}\mathcal{)}}\mathcal{\rightarrow L}^{(+)}\mathcal{(K}_{\gamma
_{1}}\mathcal{)},\text{ \ \ \ \ \ \ \ \ }A(\Omega \times \mathcal{P}_{1}(%
\mathcal{K}_{\gamma _{2}}\mathcal{)})=I_{\mathcal{K}_{\gamma _{1}}},\text{ \
\ }\forall \gamma _{1},\gamma _{2}\in \Gamma ,\text{\ }
\end{equation*}
on any outcome space $(\Omega ,\mathcal{F}_{\Omega }),$ allowed under
quantum measurements.

\subsection{Reduced quantum probabilistic model}

According to our discussion in section 5, for experiments, represented by
only linear extended generalized observables, we can equivalently replace
the probabilistic model by its reduced version.

From our consideration in section 6.1.2, it follows that, in the quantum
case, the reduced probabilistic model is given by 
\begin{equation*}
(\frak{R}_{\mathcal{S}_{q}},\mathcal{A}_{\mathcal{S}_{q}}^{ext}).
\end{equation*}
with 
\begin{equation*}
\frak{R}_{\mathcal{S}_{q}}=\{\mathcal{R}_{\mathcal{K}_{\gamma }}:\gamma \in
\Gamma \}
\end{equation*}
being the family of the sets of density operators on corresponding Hilbert
spaces and $\mathcal{A}_{\mathcal{S}_{q}}^{ext}$ being the family of all
non-trivial extended normalized $\sigma $-additive positive operator valued
measures, defined in the point (b), section 6.1.2.

\subsubsection{Quantum state instrument}

Consider a non-destructive quantum measurement $\mathcal{E}\in \lbrack
\Upsilon _{lin}^{(q)}],$ with an outcome space $(\Omega ,\mathcal{F}_{\Omega
}),$ upon a quantum system $\mathcal{S}_{q}$, described initially in terms
of a Hilbert space $\mathcal{H}.$

Suppose that, immediately after this experiment, the information on $%
\mathcal{S}_{q}$ is described in terms if a separable complex Hilbert space $%
\mathcal{K}$. Let 
\begin{equation*}
A:\mathcal{F}_{\Omega }\otimes \mathcal{B}_{\mathcal{P}_{1}(\mathcal{K)}%
}\rightarrow \mathcal{L}^{(+)}\mathcal{(H)},\text{ \ \ \ \ \ \ }A(\Omega
\times \mathcal{P}_{1}(\mathcal{K)})=I_{\mathcal{H}},
\end{equation*}
be the non-trivial extended normalized $\sigma $-additive positive operator
valued measure, which is put, due to the relation 
\begin{equation}
(\Upsilon _{lin}^{(q)}(\cdot ))(T)=\mathrm{tr}\{TA_{\Upsilon
_{lin}^{(q)}}(\cdot )\},\text{ \ \ \ \ \ \ }\forall T\in \mathcal{T(H)},
\label{61}
\end{equation}
into one-to-one correspondence with an extended linear generalized
observable $\Upsilon _{lin}^{(q)}:\mathcal{F}_{\Omega }\otimes \mathcal{B}_{%
\mathcal{P}_{1}(\mathcal{K)}}\rightarrow \mathcal{T}_{+}^{\ast }\mathcal{(H)}
$ of $\mathcal{S}_{q}.$

According to our consideration in section 4.5, the complete statistical
description of a quantum measurement $\mathcal{E}\in \lbrack \Upsilon
_{lin}^{(q)}]$ is given by the notion of a mean information state instrument 
\begin{equation*}
(\mathcal{M}_{\Upsilon _{lin}^{(q)}}^{st}(\cdot ))(\cdot ):\mathcal{F}%
_{\Omega }\times \mathcal{R}_{\mathcal{H}}\rightarrow \mathcal{T(K)}
\end{equation*}
defined, in the most general settings, by (\ref{56}).

In the quantum case, the mapping $\mathcal{M}_{\Upsilon _{lin}^{(q)}}^{st}$
is uniquely extended to all of $\mathcal{F}_{\Omega }\times \mathcal{T(H)}.$

\begin{definition}[Quantum state instrument]
\textit{We call the mapping} 
\begin{equation*}
(\mathcal{M}_{\Upsilon _{lin}^{(q)}}^{st}(\cdot ))(\cdot ):\text{ \ \ }%
\mathcal{F}_{\Omega }\times \mathcal{T(H)}\rightarrow \mathcal{T(K)}
\end{equation*}
\textit{defined by a non-trivial quantum extended linear generalized
observable }$\Upsilon _{lin}^{(q)}$ through \textit{the relation} 
\begin{equation}
(\mathcal{M}_{\Upsilon _{lin}^{(q)}}^{st}(B))(T)=\int_{\mathcal{P}_{1}(%
\mathcal{K)}}p^{\prime }(\Upsilon _{lin}^{(q)}(B\times dp^{\prime }))(T),
\label{62}
\end{equation}
for all $B\in \mathcal{F}_{\Omega },$ $T\in \mathcal{T(H)}$, \textit{a} 
\textbf{quantum state instrument}.
\end{definition}

From (\ref{62}) it follows that, for a \textit{quantum state instrument},
the mapping $\mathcal{M}_{\Upsilon _{lin}^{(q)}}^{st}(\cdot )$\ is a $\sigma 
$-additive\footnote{%
In the strong operator topology in the Banach space $\mathcal{L(T(H)},%
\mathcal{T(K))}$ of bounded linear mappings $\mathcal{T(H)}\rightarrow 
\mathcal{T(K)}.$} measure on ($\Omega ,\mathcal{F}_{\Omega })$ with values
that are\textit{\ non-negative bounded linear mappings} $\mathcal{T(H)}%
\rightarrow \mathcal{T(K)}$.

From (\ref{58}) it follows also that, for each experimental situation $%
\mathcal{E}+\mathcal{S}(\rho _{in}),$ $\rho _{in}\in \mathcal{R}_{\mathcal{H}%
},$ the probability distribution $\mu _{\Upsilon _{lin}^{(q)}}(\cdot ;\rho
_{in})$ of outcomes in $(\Omega ,\mathcal{F}_{\Omega })$ satisfies the
relation$\mathit{\ }$%
\begin{equation*}
\mu _{\Upsilon _{lin}^{(q)}}(B;\rho _{in})=\mathrm{tr}\{(\mathcal{M}%
_{\Upsilon _{lin}^{(q)}}^{st}(B))(\rho _{in})\},\text{ \ \ }\forall B\in 
\mathcal{F}_{\Omega }.
\end{equation*}

We have the following statement.

\begin{theorem}
\textit{For} \textbf{a}\textit{\ }\textbf{quantum state instrument} $%
\mathcal{M}_{\Upsilon _{lin}^{(q)}}^{(st)},$ \textit{defined} \textit{by} (%
\ref{62}):\medskip \newline
(i)\textit{\ }$\mathrm{tr}\{(\mathcal{M}_{\Upsilon _{lin}^{(q)}}^{st}(\Omega
))(T)\}=\mathrm{tr}\{T\}$, $\forall T\in \mathcal{T(H)},$ \textit{that is,
the mapping} $\mathcal{M}_{\Upsilon _{lin}^{(q)}}^{st}(\Omega )$ \textbf{is}%
\textit{\ \textbf{a} }\textbf{dynamical map};\newline
(ii)\textit{\ each value} $\mathcal{M}_{\Upsilon _{lin}^{(q)}}^{st}(B),$ $%
B\in \mathcal{F}_{\Omega },$ \textit{of the measure} 
\begin{equation*}
\mathcal{M}_{\Upsilon _{lin}^{(q)}}^{(st)}(\cdot ):\mathcal{F}_{\Omega
}\rightarrow \mathcal{L}^{(+)}\mathcal{(T(H)},\mathcal{T(K))}
\end{equation*}
\textbf{is}\textit{\ a \textbf{completely positive} bounded linear mapping }$%
\mathcal{T(H)}\rightarrow \mathcal{T(K)}.$
\end{theorem}

\begin{proof}
Due to (\ref{61}) and (\ref{62}), 
\begin{equation}
(\mathcal{M}_{\Upsilon _{lin}^{(q)}}^{st}(B))(T)=\int_{\mathcal{P}_{1}(%
\mathcal{K)}}p^{\prime }\mathrm{tr}\{TA(B\times dp^{\prime })\},\text{ \ \ \ 
}\forall T\in \mathcal{T(H)},\text{ \ \ }\forall B\in \mathcal{F}_{\Omega }.
\label{63}
\end{equation}
The relation in (i) follows trivially from (\ref{63}). In the language,
accepted in mathematical physics literature\footnote{%
See, for example, in [16].}, this means that the mapping $\mathcal{M}%
_{\Upsilon _{lin}^{(q)}}^{st}(\Omega )\in \mathcal{L}^{(+)}\mathcal{(T(H)},%
\mathcal{T(K))}$ is a dynamical map.\newline
To prove (ii), recall that a linear mapping $W:\mathcal{T(H)}\rightarrow 
\mathcal{T(K)}$ is called completely positive if, for any finite families of
elements\emph{\ }$\{f_{i}\in \mathcal{K}:i=1,...,N<\infty \}$ and $%
\{T_{i}\in \mathcal{T(H)}:i=1,...,N<\infty \}$, the sum 
\begin{equation*}
\sum_{i,j}\langle f_{i},W(T_{i}^{\ast }T_{j})f_{j}\rangle _{_{\mathcal{K}%
}}\geq 0.
\end{equation*}
Due to (\ref{63}), for any $B\in \mathcal{F}_{\Omega },$ we have 
\begin{eqnarray}
\sum_{i,j}\langle f_{i},(\mathcal{M}_{\Upsilon
_{lin}^{(q)}}^{st}(B))(T_{i}^{\ast }T_{j}))f_{j}\rangle _{_{\mathcal{K}}}
&=&\int_{\mathcal{P}_{1}(\mathcal{K})}\sum_{i,j}\langle f_{i},p^{\prime
}f_{j}\rangle _{_{\mathcal{K}}}\mathrm{tr}\{T_{i}^{\ast }T_{j}A(B\times
dp^{\prime })\}  \label{64} \\
&=&\int_{\mathcal{O}_{\mathcal{K}}}\mathrm{tr}\{G(\psi ^{\prime })A^{\prime
}(B\times d\psi ^{\prime })\},  \notag
\end{eqnarray}
where $A^{\prime }$ is the normalized $\sigma $-additive positive operator
valued measure 
\begin{eqnarray*}
A^{\prime } &:&\mathcal{F}_{\Omega }\times \mathcal{F}_{\mathcal{O}_{%
\mathcal{K}}}\rightarrow \mathcal{L}^{(+)}\mathcal{(H)},\text{ \ \ \ \ }%
A^{\prime }(\Omega \times \mathcal{O}_{\mathcal{K}})=I_{\mathcal{H}}, \\
\mathcal{F}_{\mathcal{O}_{\mathcal{K}}} &=&\{R^{-1}(F):F\in \mathcal{F}_{%
\mathcal{P}_{1}\mathcal{(K)}}\},
\end{eqnarray*}
defined by the equation 
\begin{equation*}
A^{\prime }(B\times R^{-1}(F))=A(B\times F),\text{ \ \ \ }\forall F\in 
\mathcal{B}_{\mathcal{P}_{1}(\mathcal{K)}},\text{ \ \ }\forall B\in \mathcal{%
F}_{\Omega },
\end{equation*}
and, for any $\psi ^{\prime }\in \mathcal{O}_{\mathcal{K}}$, 
\begin{equation*}
G(\psi ^{\prime })=\sum_{ji}T_{i}^{\ast }T_{j}\langle f_{i}^{\prime },\psi
^{\prime }\rangle _{_{\mathcal{K}}}\langle \psi ^{\prime },f_{j}\rangle _{_{%
\mathcal{K}}}\in \mathcal{T}^{(+)}\mathcal{(H)}.
\end{equation*}
Since $G(\psi ^{\prime })\geq 0,$ $\forall \psi ^{\prime }\in \mathcal{O}_{%
\mathcal{K}},$ in (\ref{64}) we finally have: 
\begin{equation*}
\sum_{i,j}\langle f_{i},(\mathcal{M}_{\Upsilon
_{lin}^{(q)}}^{st}(B))(T_{i}^{\ast }T_{j})f_{j}\rangle _{_{\mathcal{K}}}\geq
0,\mathcal{\ \ \ \ }\ \forall B\in \mathcal{\mathcal{F}_{\Omega }}.
\end{equation*}
\end{proof}

We underline that, in contrast to our general framework, in the frame of the
operational approach to the description of quantum measurements, \textit{the
relation between an instrument and conditional posterior quantum states, as
well as the complete positivity of a quantum instrument, are always
introduced axiomatically rather than actually proved as in this paper.}

The basics of the quantum stochastic approach to the description of quantum
measurements, formulated in [17-19], correspond to the general framework,
formulated in this paper.

\subsection{Quantum statistical model}

From our presentation in sections 5 and 6.1 it follows that, for a quantum
system $\mathcal{S}_{q},$ described initially in terms of $\mathcal{H},$ the
quantum statistical model is given by 
\begin{equation*}
(\mathcal{U}_{\mathcal{H}},\frak{M}_{\mathcal{H}}^{outcome})
\end{equation*}
where 
\begin{equation*}
\mathcal{U}_{\mathcal{H}}=(\mathcal{P}_{1}(\mathcal{H}),\mathcal{B}_{%
\mathcal{P}_{1}\mathcal{(H)}},[\mathcal{V}_{(\mathcal{P}_{1}(\mathcal{H)},%
\mathcal{B}_{\mathcal{P}_{1}\mathcal{(H)}})}])
\end{equation*}
is an initial information state space of $\mathcal{S}_{q}$ and $\frak{M}_{%
\mathcal{H}}^{outcome}$ is the family of all non-trivial POV measures: 
\begin{equation}
M:\mathcal{F}_{\Omega }\rightarrow \mathcal{L}^{(+)}\mathcal{(H)},\text{ \ \ 
}M(\Omega )=I_{\mathcal{H}},  \label{65}
\end{equation}
with any outcome space $(\Omega ,\mathcal{F}_{\Omega }).$

Correspondingly, the reduced statistical model for the description of
quantum measurements upon a system $\mathcal{S}_{q}$ is given by the pair 
\begin{equation*}
(\mathcal{R}_{\mathcal{H}},\frak{M}_{q}^{outcome}),
\end{equation*}
where $\mathcal{R}_{\mathcal{H}}$ is the set of all density operators on $%
\mathcal{H}$.

The above reduced statistical model coincides with the well-known quantum
statistical model, considered in [6,8,9,11,12,14-16,21].

\subsubsection{''No-go'' theorem}

The problem of the relation between Kolmogorov's model and the statistical
model of quantum theory is the point of intensive discussions for already
more than 70 years.

The so-called ''no-go'' theorems\footnote{%
See, for example, duscussion and references in [14,16].} state that the
properties of quantum observables can not be explained in terms of some
''underlying'' Kolmogorov's probability space.

However, different mathematical formulations of ''no-go'' theorems still
leave a loophole for doubts. The numerous papers on Bell-like inequalities
are a good confirmation that these doubts still exist.

In our framework, it is clear that, in the quantum probabilistic model,
formulated in section 6, there are no beables. In contrast to a beable, 
\textit{any quantum generalized observable represents an experiment which,
in general, perturbs a quantum system state.}

However, on the level of statistical models, where the concept of a
posterior state\ does not appear, this point is not obvious.

Based on the concept of \textit{reducible} models, which we introduced in
section 5.1, we proceed to formulate and to prove the theorem on
irreducibility of the quantum statistical model to Kolmogorov's statistical
model.

We note that, among different versions of ''no-go''theorems, the
mathematical setting of our theorem is most general and includes the
mathematical settings of all other ''no-go'' theorems\footnote{%
(See in [16], section 1.4.1.} as particular cases.

\begin{theorem}
\textit{The quantum statistical model is} \textbf{not reducible} \textit{to
Kolmogorov's statistical model}.\smallskip
\end{theorem}

\begin{proof}
We are based on the definitions, in sections 5.1, 5.2, 6.3, of the notions
of reducible models, Kolmogorov's statistical model and the quantum
statistical model, respectively.\newline
Suppose\ that the quantum statistical model is reducible to Kolmogorov's
statistical model. \newline
Then there must exist an information state space ($\Theta _{in},\mathcal{F}%
_{\Theta _{in}},[\mathcal{V}_{(\Theta _{in},\mathcal{F}_{\Theta _{in}})}])$
such that:\newline
(a) the quantum information state space $(\mathcal{P}_{1}(\mathcal{H}),%
\mathcal{B}_{\mathcal{P}_{1}\mathcal{(H)}},[\mathcal{V}_{(\mathcal{P}_{1}(%
\mathcal{H)},\mathcal{B}_{\mathcal{P}_{1}\mathcal{(H)}})}])$ is induced by
an information state space $(\Theta _{in},\mathcal{F}_{\Theta _{in}},[%
\mathcal{V}_{(\Theta _{in},\mathcal{F}_{\Theta _{in}})}]);$\newline
(b) any quantum outcome generalized observable $\mathrm{M}_{lin}^{(q)}$ on $(%
\mathcal{P}_{1}(\mathcal{H}),\mathcal{B}_{\mathcal{P}_{1}\mathcal{(H)}})$
and any beable $\mathrm{S}^{(be)}$ on $(\Theta _{in},\mathcal{F}_{\Theta
_{in}})$, with the outcome space $(\mathcal{P}_{1}(\mathcal{H}),\mathcal{B}_{%
\mathcal{P}_{1}\mathcal{(H)}}),$ generate, due to (\ref{22}), a beable 
\begin{equation*}
\mathrm{M}^{(be)}(B)=(\mathrm{M}_{lin}^{(q)}(B)\circ \mathrm{S}^{(be)}),%
\text{ \ \ \ }\forall B\in \mathcal{F}_{\Omega },
\end{equation*}
on $(\Theta _{in},\mathcal{F}_{\Theta _{in}}).$ \newline
Since, due to (\ref{60}), 
\begin{equation*}
(\mathrm{M}_{lin}^{(q)}(B))(p)=\mathrm{tr}\{pM(B)\},\text{ \ \ }\forall p\in 
\mathcal{P}_{1}(\mathcal{H)}\text{, \ }\forall B\in \mathcal{F}_{\Omega },
\end{equation*}
where $M$ is a POV measure$,$ we derive: 
\begin{equation}
(\mathrm{M}^{(be)}(B))(\theta _{in})=\int_{\mathcal{P}_{1}(\mathcal{H)}}%
\mathrm{tr}\{pM(B)\}(\mathrm{S}^{(be)}(dp))(\theta _{in}),  \label{66}
\end{equation}
for all $B\in \mathcal{F}_{\Omega },$ $\theta _{in}\in \Theta _{in}.$\newline
Due to the definition of Kolmogorov's model, section 5.2, and definition 14
of a beable, there must exist:\newline
a probability state space $(\Theta ,\mathcal{F}_{\Theta },[\mathcal{V}%
_{(\Theta ,\mathcal{F}_{\Theta })}])$; \newline
an $\mathcal{F}_{\Theta }/\mathcal{F}_{\Theta _{in}}$-measurable mapping $%
\phi _{in}:\Theta \rightarrow \Theta _{in};$\newline
an $\mathcal{F}_{\Theta }/\mathcal{F}_{\Omega }$-measurable mapping $\varphi
:\Theta \rightarrow \Omega ;$\newline
an $\mathcal{F}_{\Theta }/\mathcal{B}_{\mathcal{P}_{1}\mathcal{(H)}}$%
-measurable mapping $\Phi :\Theta \rightarrow \mathcal{P}_{1}(\mathcal{H)};$%
\newline
such that 
\begin{eqnarray}
(\mathrm{E}_{\phi _{in}^{-1}}^{(be)}(B))(\theta ) &\equiv &(\mathrm{E}%
^{(be)}(B))(\phi _{in}(\theta ))=\chi _{\varphi ^{-1}(B)}(\theta ),\text{ \
\ \ }\forall B\in \mathcal{F}_{\Omega },\text{ \ \ \ }\forall \theta \in
\Theta ,  \label{67} \\
(\mathrm{S}_{\phi _{in}^{-1}}^{(be)}(F))(\theta ) &\equiv &(\mathrm{S}%
^{(be)}(F))(\phi _{in}(\theta ))=\chi _{\Phi ^{-1}(F)}(\theta ),\text{ \ \ \ 
}\forall F\in \mathcal{B}_{\mathcal{P}_{1}\mathcal{(H)}},\text{ \ }\forall
\theta \in \Theta .  \notag
\end{eqnarray}
Taking $\phi _{in}$-preimages of the left and the right hand sides in (\ref
{66}) and considering (\ref{67}), we finally derive that the assumptions
(a), (b) result in the following relation: 
\begin{equation}
\chi _{\varphi ^{-1}(B)}(\theta )=\mathrm{tr}\{p_{\theta }M(B)\},\text{ \ \
\ }p_{\theta }=\Phi (\theta )\in \mathcal{P}_{1}(\mathcal{H)},\text{ \ \ }%
\forall B\in \mathcal{F}_{\Omega },\text{ \ \ \ }\forall \theta \in \Theta .
\label{68}
\end{equation}
Take, for example, a discrete projection valued measure on $(\mathbb{R},%
\mathcal{B(}\mathbb{R))}:$ 
\begin{equation*}
P(B)=\sum_{\lambda _{i}\in B}|\psi _{i}\rangle \langle \psi _{i}|,\text{ \ \
\ }\lambda _{i}\in \mathbb{R},\text{ \ }\langle \psi _{i},\psi _{j}\rangle _{%
\mathcal{H}}=\delta _{ij},\text{ \ }i,j=1,2,...,N\leq \infty ;\text{\ \ }%
\forall B\in \mathcal{B(}\mathbb{R)}.
\end{equation*}
In the most general settings, for any $\theta \in \Theta ,$ the value $%
p_{\theta }=\Phi (\theta )\in \mathcal{P}_{1}(\mathcal{H)}$ admits the
representation: 
\begin{equation}
p_{\theta }=|\psi _{\theta }\rangle \langle \psi _{\theta }|,\text{ \ \ \ \ }%
\psi _{\theta }=\sum_{i}\beta _{i}^{(\theta )}\psi _{i},\text{ \ \ \ }%
\sum_{i}|\beta _{i}^{(\theta )}|^{2}=1,  \label{69}
\end{equation}
where, to a given $p_{\theta },$ a vector $\psi _{\theta }\in \mathcal{O}_{%
\mathcal{H}}$ is defined up to phase equivalence. \newline
Substituting (\ref{69}) into (\ref{68}), we derive: 
\begin{equation*}
\chi _{\varphi ^{-1}(B)}(\theta )=\sum_{\lambda _{i}\in B}|\beta
_{i}^{(\theta )}|^{2},\text{ \ \ }\forall B\in \mathcal{B(}\mathbb{R)},\text{
\ \ }\forall \theta \in \Theta .
\end{equation*}
But this relation cannot be valid for all $\theta \in \Theta $ and all $B\in 
\mathcal{B(}\mathbb{R)}.$ \newline
Thus, assuming (a), (b), we have come to the contradiction and, hence, the
quantum statistical (and, hence, probabilistic) model is not \textit{%
reducible }to Kolmogorov's statistical model.
\end{proof}

\begin{conclusion}
Kolmogorov's model cannot induce the properties of quantum generalized
observables.
\end{conclusion}

\section{Concluding remarks}

In the present paper we formulate a new general framework for the
probabilistic description of an experiment upon a system in terms of initial
information representing this system.

We introduce the notions of an information state, an information state space
and a generalized observable and prove the corresponding statements.

We prove that, to any experiment upon a system, there corresponds a unique
generalized observable on a system initial information state space. An
initial information state space provides the knowledge on the description of
only such experiments which are represented on this space by non-trivial
generalized observables.

We specify the special types of generalized observables:

(i) observables, which describe experiments with deterministic set-up;

(ii) beables, which describe non-perturbing experiments with deterministic
set-up.

A beable describes\ a non-perturbing experiment on the ''errorless''
measurement of some property of a system, objectively existing before this
experiment. Under an experiment, described by a beable, the randomness is
caused only by the uncertainty encoded in the initial probability
distribution in a Kolmogorov probability space. In Kolmogorov's model only
beables are considered.

In general, however, a generalized observable represents an experiment upon
a system in a non-separable manner and cannot be associated with any of
system properties, objectively existing before this experiment.

In the most general settings, we introduce the concept of a complete
information description of a non-destructive experiment upon a system. This
type of description is given by the notion of an information state
instrument, which we define in this paper.

We point out that the phenomenon of ''reduction'' of an information state is
inherent, in general, to any non-destructive experiment and upon a system of
any type. In case of non-destructive experiments upon quantum systems, the
von Neumann ''state collapse'' and its further generalizations represent
particular cases of this general phenomenon.

For a system, described by an information state space with a Banach space
based structure, we introduce, in the most general settings, the notion of a
mean information state instrument.

We specify the concepts of the probabilistic model and the statistical model
for the description of experiments upon a quantum system and prove, in the
most general mathematical setting, the theorem on the irreducibility of the
quantum statistical model to Kolmogorov's statistical model.

In the quantum case, only such a generalized observable is allowed which is
represented by a $\sigma $-additive measure with values that are
non-negative continuous linear\textit{\ }functionals on the Banach space of
trace class operators. To each quantum generalized observable there is put
into the one-to-one correspondence a normalized $\sigma $-additive measure
with values that are non-negative bounded linear operators on a Hilbert
space. A projection valued measure represents a quantum measurement with a
deterministic set-up.

We prove that a quantum state instrument represents a dynamical map and is
completely positive.

\paragraph{Acknowledgements}

\noindent I am indebted to Ole E. Barndorff-Nielsen for valuable remarks. I
am also thankful to Goran Peskir for stimulating discussions.

The support, given by MaPhySto, for the research, reported here and
elsewhere, is gratefully acknowledged.

\bigskip

\bigskip

MaPhySto

Department of Mathematical Sciences

University of Aarhus

Ny Munkegade \ DK-8000 \ Aarhus C

Denmark\medskip

e-mail: \ elena@imf.au.dk

\ \ \ \ \ \ \ \ \ \ \ erl@erl.msk.ru

\end{document}